\documentclass[12pt]{article}
\begin{document}
\newcommand{\beq}{\begin{equation}}
\newcommand{\eeq}{\end{equation}}
\newcommand{\beqa}{\begin{eqnarray}}
\newcommand{\eeqa}{\end{eqnarray}}
\newcommand{\beqar}{\begin{eqnarray*}}
\newcommand{\eeqar}{\end{eqnarray*}}
\newcommand{\al}{\alpha}
\newcommand{\be}{\beta}
\newcommand{\del}{\delta}
\newcommand{\D}{\Delta}
\newcommand{\eps}{\epsilon}
\newcommand{\ga}{\gamma}
\newcommand{\Ga}{\Gamma}
\newcommand{\ka}{\kappa}
\newcommand{\inn}{\!\cdot\!}
\newcommand{\h}{\eta}
\newcommand{\kk}{\varphi}
\newcommand\F{{}_3F_2}
\newcommand{\la}{\lambda}
\newcommand{\La}{\Lambda}
\newcommand{\na}{\nabla}
\newcommand{\Om}{\Omega}
\newcommand{\p}{\phi}
\newcommand{\sig}{\sigma}
\renewcommand{\t}{\theta}
\newcommand{\z}{\zeta}
\newcommand{\ssc}{\scriptscriptstyle}
\newcommand{\eg}{{\it e.g.,}\ }
\newcommand{\ie}{{\it i.e.,}\ }
\newcommand{\labell}[1]{\label{#1}} 
\newcommand{\reef}[1]{(\ref{#1})}
\newcommand\prt{\partial}
\newcommand\veps{\varepsilon}
\newcommand\ls{\ell_s}
\newcommand\cF{{\cal F}}
\newcommand\cA{{\cal A}}
\newcommand\cS{{\cal S}}
\newcommand\cT{{\cal T}}
\newcommand\cC{{\cal C}}
\newcommand\cL{{\cal L}}
\newcommand\cM{{\cal M}}
\newcommand\cN{{\cal N}}
\newcommand\cG{{\cal G}}
\newcommand\cI{{\cal I}}
\newcommand\cJ{{\cal J}}
\newcommand\cl{{\iota}}
\newcommand\cP{{\cal P}}
\newcommand\cQ{{\cal Q}}
\newcommand\cg{{\it g}}
\newcommand\cR{{\cal R}}
\newcommand\cB{{\cal B}}
\newcommand\cO{{\cal O}}
\newcommand\tcO{{\tilde {{\cal O}}}}
\newcommand\bz{\bar{z}}
\newcommand\bw{\bar{w}}
\newcommand\hF{\hat{F}}
\newcommand\hA{\hat{A}}
\newcommand\hT{\hat{T}}
\newcommand\htau{\hat{\tau}}
\newcommand\hD{\hat{D}}
\newcommand\hf{\hat{f}}
\newcommand\hg{\hat{g}}
\newcommand\hp{\hat{\phi}}
\newcommand\hi{\hat{i}}
\newcommand\ha{\hat{a}}
\newcommand\hQ{\hat{Q}}
\newcommand\hP{\hat{\Phi}}
\newcommand\hS{\hat{S}}
\newcommand\hX{\hat{X}}
\newcommand\tL{\tilde{\cal L}}
\newcommand\hL{\hat{\cal L}}
\newcommand\tG{{\widetilde G}}
\newcommand\tg{{\widetilde g}}
\newcommand\tphi{{\widetilde \phi}}
\newcommand\tPhi{{\widetilde \Phi}}
\newcommand\te{{\tilde e}}
\newcommand\tk{{\tilde k}}
\newcommand\tf{{\tilde f}}
\newcommand\tF{{\widetilde F}}
\newcommand\tK{{\widetilde K}}
\newcommand\tE{{\widetilde E}}
\newcommand\tpsi{{\tilde \psi}}
\newcommand\tX{{\widetilde X}}
\newcommand\tD{{\widetilde D}}
\newcommand\tO{{\widetilde O}}
\newcommand\tS{{\tilde S}}
\newcommand\tB{{\widetilde B}}
\newcommand\tA{{\widetilde A}}
\newcommand\tT{{\widetilde T}}
\newcommand\tC{{\widetilde C}}
\newcommand\tV{{\widetilde V}}
\newcommand\thF{{\widetilde {\hat {F}}}}
\newcommand\Tr{{\rm Tr}}
\newcommand\tr{{\rm tr}}
\newcommand\STr{{\rm STr}}
\newcommand\M[2]{M^{#1}{}_{#2}}
\parskip 0.3cm

\vspace*{1cm}

\begin{center}
{\bf \Large  Ramond-Ramond S-matrix elements \\ from  T-dual  Ward identity   }

\vspace*{1cm}

{Komeil Babaei Velni\footnote{komeilvelni@gmail.com} and Mohammad R. Garousi\footnote{garousi@um.ac.ir} }\\
\vspace*{1cm}
{ Department of Physics, Ferdowsi University of Mashhad,\\ P.O. Box 1436, Mashhad, Iran}
\\
\vspace{2cm}

\end{center}

\begin{abstract}
\baselineskip=18pt

Recently it has been  speculated  that  the Ward identities associated with the string dualities and the gauge symmetries can be used as  guiding principles to find all components of the scattering amplitude of $n$ supergravitons from a given component of the S-matrix. In this paper, we apply the Ward identities associated with the T-duality and the gauge symmetries  on the  disk-level  S-matrix element of one  RR $(p-3)$-form, one   NSNS and one  NS   states,  to find the corresponding S-matrix elements of the RR $(p-1)$-form, $(p+1)$-form or the RR$(p+3)$-form  on the world volume of a D$_p$-brane. Moreover, we apply these Ward identities on the   S-matrix element of one RR $(p-3)$-form  and two  NSNS   states  to find the corresponding  S-matrix elements of  the RR $(p-1)$-form, $(p+1)$-form, $(p+3)$-form or the  RR $(p+5)$-form.  

\end{abstract}
Keywords:   T-duality, Ward identity, Chern-Simons  couplings

\setcounter{page}{0}
\setcounter{footnote}{0}
\newpage




\section{Introduction } \label{intro}
Higher-derivative couplings in superstring theory can be captured from $\alpha'$-expansion of the  corresponding  S-matrix elements \cite{Gross:1986iv,Gross:1986mw} and from exploring the   dualities of the superstring theory \cite{Kikkawa:1984cp}-\cite{Hull:1994ys}.    The dualities can be implemented either on-shell or off-shell. At the on-shell level, they appear in the S-matrix elements as S-dual and  T-dual Ward identities \cite{Garousi:2011we}-\cite{Velni:2012sv}. These identities establish connections between  different elements of the scattering amplitude of  $n$ supergravitons. Calculating one  element explicitly in the world sheet conformal field theory, then all other elements of the S-matrix may be found by the Ward identities.  At the off-shell level, on the other hand, the dualities appear as symmetries of the effective action. Calculating the couplings of one specific component of the supergraviton at order $\alpha'^n$ from the corresponding S-matrix element, then the couplings of all other components  at this order may be found by  the dualities \cite{Garousi:2009dj}-\cite{Garousi:2013qka}.

The effective actions of D$_p$-branes in superstring theory at leading order of $\alpha'$ are given by the Dirac-Born-Infeld (DBI) and the Chern-Simons (CS) actions which are invariant under off-shell  T-duality. The first higher-derivative correction to these actions is at order $\alpha'^2$. The curvature squared corrections to the DBI action has been found in \cite{Bachas:1999um} from the $\alpha'$-expansion of the disk-level S-matrix element of two gravitons \cite{Garousi:1996ad}. At the $\alpha'^2$-order, the  S-matrix element of  two massless closed string states in the superstring theory has only contact terms, \eg for two gravitons, they are the curvature squared couplings in the momentum space \cite{Bachas:1999um}. The T-dual and S-dual Ward identities then dictate that the curvature couplings must be invariant under linear T-duality and S-duality. The consistency of the curvature couplings with the linear T-duality and S-duality has been used in \cite{Garousi:2009dj,Garousi:2011fc} to find the on-shell couplings of two supergravitons  on the world-volume of D$_p$-brane at order $\alpha'^2$.   

The curvature  corrections to the CS action, on the other hand,  has been first found in \cite{Green:1996dd,Cheung:1997az,Minasian:1997mm}  by requiring that the chiral anomaly on the world volume of intersecting D-branes (I-brane) cancels with the anomalous variation of the CS action.  These corrections also starts at order $\alpha'^2$ which is the  curvature squared times the RR potential, \ie $C^{(p-3)}\wedge R\wedge R$. These couplings have been confirmed  in \cite{Craps:1998fn,Morales:1998ux,Stefanski:1998yx} by the $\alpha'$-expansion of the disk-level S-matrix element of two gravitons and one RR vertex operator. At the $\alpha'^2$-order,  the S-matrix element has only contact terms which are the coupling $C^{(p-3)}\wedge R\wedge R$ in the momentum space. However,  all other S-matrix elements of three massless closed strings have both contact terms and massless poles. As a result, the T-dual   ward identity does not indicate that the curvature couplings must be invariant under the linear T-duality.  On the other hand, it has been observed in \cite{Garousi:2010ki} that the CS action at order $\alpha'^2$ has also couplings between one NSNS and one RR $n$-form where $n=p-1,\, p+1,\, p+3$.  We expect the combination of these couplings and the curvature squared couplings  should be extendible to the  off-shell nonlinear T-duality after including many other couplings at order $\alpha'^2$. Some of these D$_p$-brane couplings   involving the RR $(p-3)$-form  have been found in \cite{Garousi:2011ut,Becker:2011ar} from the $\alpha'$ expansion of the corresponding S-matrix elements.

The couplings involving the RR $(p-3)$-form reveal that it is a hard task to find all  other couplings involving  the   RR $n$-form where $n=p-1,\, p+1,\, p+3$, from the off-shell nonlinear T-duality requirement.  Partial results for such couplings, however,  have been found in \cite{Becker:2010ij,Garousi:2010rn}. We are interested in finding such couplings from the $\alpha'$ expansion of the corresponding S-matrix elements. We are going to benefit from  the on-shell linear T-duality requirement to find the S-matrix elements from the S-matrix elements of the RR $(p-3)$-form which have been calculated explicitly in \cite{Garousi:2011ut,Becker:2011ar}. The implicit assumption  in the T-duality transformation  that fields must be independent of the Killing coordinate,   causes in some cases that the T-dual Ward identity not to be able to capture the  new  S-matrix elements   in all details. However, the Ward identity corresponding to the gauge symmetries  can be used to fix this problem \cite{Velni:2012sv}.


The disk-level scattering amplitude of one massless RR $(p-3)$-form, one NSNS state and one  open string  NS state   has been calculated  in \cite{Becker:2011ar,Garousi:2012gh}. The RR potential in this amplitude   carries either one or zero transverse index. Accordingly it can be split into two parts. The amplitude corresponding to the first part has one integral representing the closed and open string channels. The amplitude corresponding to  the second part has three integrals which satisfy one constraint equation. 
The T-dual Ward identity connects the amplitude of the RR $(p-3)$-form to the amplitudes of the RR $(p-1)$-form, $(p+1)$-form and the RR $(p+3)$-form which furnish a T-dual multiplet. The T-dual multiplet corresponding to the first part   has been found in \cite{Velni:2012sv}. It has the following structure:
\beqa
 {A}_1(C_i^{(p-3)})\rightarrow {A}_2(C_{ij}^{(p-1)})\rightarrow {A}_3(C_{ijk}^{(p+1)})\rightarrow {A}_4(C_{ijkl}^{(p+3)})\labell{Bi}
 \eeqa
where the number in the label of $A$ refers to the number of transverse indices of the RR potential. All other indices of the RR potential   contract with the world volume form.  The   components $A_2,\, A_3,\, A_4$ carry the same integral that the first component $A_1$ carries. In this case, the T-dual Ward identity captures the new S-matrix elements in full details. This is confirmed by the fact that each S-matrix element satisfies the Ward identities corresponding to the  NSNS and NS gauge transformations \cite{Velni:2012sv}. The multiplet does not satisfy the Ward identity corresponding to the RR gauge transformation because it contains only the first part of the RR $(p-3)$-form which has one transverse index. 

In this paper, we will find, among other things,  the T-dual multiplets corresponding to the second part which has the RR $(p-3)$-form with no transverse index. The result has the following structure:
\beqa
\matrix{A_0(C ^{(p-3)})& \!\!\!\!\!\rightarrow \!\!\!\!\!& A_1(C_{ i}^{(p-1)})& \!\!\!\!\!\rightarrow \!\!\!\!\!&A_2(C_{ ij}^{(p+1)})&  \!\!\!\!\!\rightarrow \!\!\!\!\! & A_3(C_{ ijk}^{(p+3)})\cr 
&&\downarrow & &\downarrow&& \downarrow\cr
&&  {A_1'} (C_{ i}^{(p-1)})& \!\!\!\!\!\rightarrow \!\!\!\!\!& {A_2'} (C_{ ij}^{(p+1)})& \!\!\!\!\!\rightarrow \!\!\!\!\!& {A_3'} (C_{ ijk}^{(p+3)})\cr} \labell{2close1open} 
\eeqa
where the horizontal arrows show the linear T-duality transformation and the vertical arrows show the NSNS or the NS gauge transformations. In this case the amplitudes in the first line which are connected by the linear  T-duality transformation, do not satisfy the Ward identity associated with the NSNS or the NS gauge transformations. The amplitudes in the second line which are connected by the T-duality, are added to make the whole amplitude to be invariant under the NSNS and the NS gauge transformations. All the above components carry the same three integrals that the first component carries. 

One may expect the sum of the multiplets \reef{Bi} and \reef{2close1open} to satisfy the Ward identity corresponding to the RR gauge transformation. This is the case for the first components which are calculated explicitly. However, as we shall show the other components which are calculated through the Ward identity corresponding to the T-duality and the NSNS gauge transformations, do not satisfy this condition. This indicates that there should be another T-dual multiplet whose first component is $A_0(C ^{(p-1)})$. This component which should be invariant under the RR Ward identity, is connected to the $C^{(p-1)}$ amplitudes in \reef{2close1open} by the Ward identity corresponding to the NSNS and the NS  gauge transformations. Imposing this condition, we  will be able to find this component. The amplitude has two new integrals which satisfy one new constraint equation. We will also find the  other amplitudes which are connected to it by the  Ward identity. The multiplet has the following structure:
\beqa
\matrix{A_0(C ^{(p-1)})& \!\!\!\!\!\rightarrow \!\!\!\!\!& A_1(C_{ i}^{(p+1)})&   \cr
&& \downarrow\cr
&&  {A_1'} (C_{ i}^{(p+1)})\cr
} \labell{2new} 
\eeqa
The other components have the same integrals that the first component has. The sum of the multiplets \reef{Bi}, \reef{2close1open} and \reef{2new} satisfies the Ward identity corresponding to all the gauge symmetries and the T-duality. 


The disk-level S-matrix element of  one  RR $(p-3)$-form and two  NSNS states    has been calculated  in \cite{Garousi:2010bm,Garousi:2011ut,Becker:2011bw,Becker:2011ar}. The RR potential in this amplitude  carries two,  one or zero transverse indices. Accordingly it has three parts. The amplitude for the first part has one integral, the amplitude for the second part has 5 integrals which satisfy  two constraint equations, and the amplitude for the third part has 14 integrals which satisfy 8 constraint equations. The T-dual Ward identity connects these three parts to the amplitudes of the RR $(p-1)$-form, $(p+1)$-form, $(p+3)$-form and the RR $(p+5)$-form. The T-dual multiplets corresponding to the first part  has been found in \cite{Velni:2012sv}. It has the following structure:
\beqa
 {A}_2(C_{ij}^{(p-3)})\rightarrow {A}_3(C_{ijk}^{(p-1)})\rightarrow {A}_4(C_{ijkl}^{(p+1)})\rightarrow {A}_5(C_{ijklm}^{(p+3)})\rightarrow {A}_6(C_{ijklmn}^{(p+5)})\labell{Ai}
 \eeqa
Each amplitude satisfies the Ward identity corresponding to the NSNS gauge transformation. They all contain one integral. The T-dual multiplet corresponding to the second part which has been found in \cite{Velni:2012sv} has the following structure:
\beqa
\matrix{\cA_1(C_i ^{(p-3)})& \!\!\!\!\!\rightarrow \!\!\!\!\!& \cA_2(C_{ij} ^{(p-1)})& \!\!\!\!\!\rightarrow \!\!\!\!\!&\cA_3(C_{ijk} ^{(p+1)})&  \!\!\!\!\!\rightarrow \!\!\!\!\! & \cA_4(C_{ijkl} ^{(p+3)})& \!\!\!\!\!\rightarrow \!\!\!\!\!&\cA_5(C_{ijklm} ^{(p+5)})\cr 
&&\downarrow & &\downarrow&& \downarrow&& \downarrow\cr
&&  \cA'_2(C_{ij} ^{(p-1)})& \!\!\!\!\!\rightarrow \!\!\!\!\!&\cA'_3(C_{ijk} ^{(p+1)})& \!\!\!\!\!\rightarrow \!\!\!\!\!&\cA'_4(C_{ijkl} ^{(p+3)})& \!\!\!\!\!\rightarrow \!\!\!\!\!&\cA'_5 (C_{ijklm} ^{(p+5)})}
 \labell{close3} 
 \eeqa 
where the T-dual multiplet in the second line is needed for the NSNS gauge symmetry.  Each component contains the same 5 integrals that the first component has. It has been speculated in \cite{Velni:2012sv} that there are three multiplets corresponding to the third part. In this paper, we will find these multiplets. We will find they have  the following structure:
\beqa
\matrix{{\bf A}_0(C ^{(p-3)})& \!\!\!\!\!\rightarrow \!\!\!\!\!& {\bf A}_1(C_i ^{(p-1)})& \!\!\!\!\!\rightarrow \!\!\!\!\!&{\bf A}_2(C_{ij} ^{(p+1)})&  \!\!\!\!\!\rightarrow \!\!\!\!\! & {\bf A}_3(C_{ijk} ^{(p+3)})& \!\!\!\!\!\rightarrow \!\!\!\!\!&{\bf A}_4(C_{ijkl} ^{(p+5)})\cr 
&&\downarrow & &\downarrow&& \downarrow&& \downarrow\cr
&&  {\bf A}'_1(C_i ^{(p-1)})& \!\!\!\!\!\rightarrow \!\!\!\!\!&{\bf A}'_2(C_{ij} ^{(p+1)})& \!\!\!\!\!\rightarrow \!\!\!\!\!&{\bf A}'_3(C_{ijk} ^{(p+3)})& \!\!\!\!\!\rightarrow \!\!\!\!\!&{\bf A}'_4(C_{ijkl} ^{(p+5)})\cr
&&&& \downarrow & &\downarrow&&   \cr
&&&&A''_2(C_{ij} ^{(p+1)})& \!\!\!\!\!\rightarrow \!\!\!\!\!&A''_3(C_{ijk} ^{(p+3)})& & }\labell{3close0}
   \eeqa  
The multiplets in the second and in the third lines are needed for the NSNS gauge symmetry. All the above components carry the same 14 integrals that the first component carries. Here also one may expect the sum of the multiplets \reef{Ai}, \reef{close3} and \reef{3close0} to satisfy the Ward identity corresponding to the RR gauge transformation. Eventhough  the first components which are calculated explicitly, satisfy this condition, the other components  do not satisfy this condition. This again indicates that there should be another T-dual multiplet like \reef{2new}  whose first component is  invariant under the RR gauge transformation. In this case we find that it is hard to find this amplitude from the Ward identities of the NSNS gauge transformations. This component may be calculated explicitly in string theory in which we are not interested in this paper.  

The outline of the paper is as follows: We begin with section 2 which is a  review for  the T-dual Ward identity.   In section 3, using the consistency of  the S-matrix element of one RR $(p-3)$-form, one NSNS state and one  open string  NS state  which has been calculated  in \cite{Becker:2011ar,Garousi:2012gh}, with the Ward identity corresponding to the T-duality and the   gauge symmetries, we find the corresponding S-matrix elements for all other RR potentials. In section 4, we perform the same calculations for the S-matrix element of one RR $(p-3)$-form and two NSNS states    which has been calculated  in \cite{ Garousi:2011ut,Becker:2011ar}. The amplitudes in this section, however, do not fully satisfy the Ward identity corresponding to the  gauge transformation because of our lack of knowledge of the amplitude of the RR field strength $F^{(p)}$ with no transverse index.  In section 5, we   briefly discuss our results.

\section{T-dual Ward identity}

It is known that the gauge symmetries of a given theory appear in the S-matrix elements through the corresponding Ward identities. That is,  the S-matrix elements of the theory should be invariant under the linearized gauge transformations on the external states and should be invariant under the full nonlinear gauge transformation on the background fields. This idea has been speculated in \cite{Garousi:2011we} to be hold even for the duality transformations of the theory. In particular, the S-matrix elements should be invariant/covariant under linear T-duality transformation of the external states and  under nonlinear T-duality transformation of the background fields.

The full set of nonlinear T-duality transformations for massless RR and NSNS fields have been found in \cite{TB,Meessen:1998qm,Bergshoeff:1995as,Bergshoeff:1996ui,Hassan:1999bv}. The nonlinear T-duality transformations of the  RR field $C$ and the antisymmetric field $B$ are such that  the expression ${\cal C}=e^BC$  transforms linearly under  T-duality \cite{Taylor:1999pr}. When  the T-duality transformation acts along the Killing coordinate $y$,  the massless NSNS fields and  ${\cal C}$  transforms as:
\beqa
e^{2\tphi}=\frac{e^{2\phi}}{G_{yy}}&;& 
\tG_{yy}=\frac{1}{G_{yy}}\nonumber\\
\tG_{\mu y}=\frac{B_{\mu y}}{G_{yy}}&;&
\tG_{\mu\nu}=G_{\mu\nu}-\frac{G_{\mu y}G_{\nu y}-B_{\mu y}B_{\nu y}}{G_{yy}}\nonumber\\
\tB_{\mu y}=\frac{G_{\mu y}}{G_{yy}}&;&
\tB_{\mu\nu}=B_{\mu\nu}-\frac{B_{\mu y}G_{\nu y}-G_{\mu y}B_{\nu y}}{G_{yy}}\nonumber\\
{\cal \tC}^{(n)}_{\mu\cdots \nu y}={\cal C}^{(n-1)}_{\mu\cdots \nu }&;&
{\cal \tC}^{(n)}_{\mu\cdots\nu}={\cal C}^{(n+1)}_{\mu\cdots\nu y}\labell{Cy}
\eeqa
where $\mu,\nu$ denote any coordinate directions other than $y$. In above transformation the metric is given in the string frame. If $y$ is identified on a circle of radius $R$, \ie $y\sim y+2\pi R$, then after T-duality the radius becomes $\tilde{R}=\alpha'/R$. The string coupling is also shifted as $\tilde{g}_s=g_s\sqrt{\alpha'}/R$.

We would like to study the T-dual Ward identity of  scattering amplitudes, so we need the above transformations at the linear order. Assuming that the NSNS  fields are small perturbations around the flat space, \eg $G_{\mu\nu}=\eta_{\mu\nu}+h_{\mu\nu}$, the above  transformations take the following linear form:
\beqa
&&
\tilde{\phi}=\phi-\frac{1}{2}h_{yy},\,\tilde{h}_{yy}=-h_{yy},\, \tilde{h}_{\mu y}=B_{\mu y},\, \tilde{B}_{\mu y}=h_{\mu y},\,\tilde{h}_{\mu\nu}=h_{\mu\nu},\,\tilde{B}_{\mu\nu}=B_{\mu\nu}\nonumber\\
&&{  \tC}^{(n)}_{\mu\cdots \nu y}={  C}^{(n-1)}_{\mu\cdots \nu },\,\,\,{  \tC}^{(n)}_{\mu\cdots\nu}={  C}^{(n+1)}_{\mu\cdots\nu y}\labell{linear}
\eeqa
The T-duality transformation of the gauge field on the world volume of D-brane, when it is along the Killing direction, is $\tilde{A}_y=\phi^y$ where $\phi^y$ is the transverse scalar. When the gauge field is not along the Killing direction it is invariant under the T-duality.

The method for finding  the couplings which are invariant under the T-duality    is given in \cite{Garousi:2009dj}. It can be used to find the T-dual multiplet corresponding to a given scattering amplitude which satisfies the T-dual Ward identity. Let us review this method here.  
Suppose we are implementing T-duality along a world volume direction $y$ of a D$_p$-brane. We first separate the  world-volume indices  along and orthogonal to $y$,     and then apply the  T-duality transformations \reef{linear}.  The  orthogonal indices  are  the complete world-volume indices   of the  T-dual D$_{p-1}$-brane. However,  $y$  in the T-dual theory, which is a normal bundle index, is not  complete.  
On the other hand, the normal bundle indices of  the original theory  are not complete in the T-dual D$_{p-1}$-brane. They do not include the $y$ index.  In a T-dual multiplet, the index  $y$   must be combined  with the incomplete normal bundle indices  to make them complete.  If the  scattering amplitudes are not invariant under the T-duality,  one should then add  new amplitudes  to them  to have   the complete indices  after the T-duality transformation. In this way one can find the T-dual multiplet which satisfies the T-dual Ward identity.

The linear T-duality transformation of the RR potential \reef{linear} reveals that the D$_p$-brane world volume couplings of the RR $n$-form which have no transverse index are not related by the T-duality to the couplings in which  the RR $n$-form have one transverse index. The couplings in which the RR $n$-form have one transverse index are not related by the T-duality to the couplings in which  the RR $n$-form have two transverse indices, and so on. To clarify this one may first write $n=p+m$. If T-duality is implemented along a world volume direction  of a D$_p$-brane, then the RR $(p+m)$-form with no transverse index transforms to the RR $(p+m+1)$-form with one transverse index, however,  at the same time the D$_p$-brane transforms to D$_{p-1}$-brane. As a result, the RR $n$-form with no transverse index does not transform to the  RR $(n)$-form with one transverse index. It transforms to  RR $(n+2)$-form with one transverse index. Therefore, to study the T-duality of the world volume amplitudes involving the RR potential, it is convenient to classify the RR potential according to its transverse indices.

\section{ Two closed and one open string amplitudes}

The disk-level S-matrix element of one RR $(p-3)$-form, one NSNS state and one open string NS state has been calculated in \cite{Becker:2011ar,Garousi:2012gh}.  The amplitude  is nonzero only for the case that the NSNS polarization tensor is antisymmetric, the open string is the gauge field and  the RR polarization tensor has one and zero transverse index. Accordingly the amplitude has two parts which should be studied under the T-dual Ward identity separately.   The first part is \footnote{Our conventions set $\alpha'=2$ in the string theory amplitudes.
Our index convention is that the Greek letters  $(\mu,\nu,\cdots)$ are  the indices of the space-time coordinates, the Latin letters $(a,d,c,\cdots)$ are the world-volume indices and the letters $(i,j,k,\cdots)$ are the normal bundle indices.}
\beqa
A_1(C_i^{(p-3)})&\sim& T_p(\veps_1 )_i{}^{a_5\cdots a_p}\eps_{a_0\cdots a_p}p_3^ip_3^{a_4}(\veps_3^A)^{a_3a_2}p_2^{a_0}\veps_2^{a_1}{\cal Q}\labell{A11}
\eeqa
where   ${\cal Q}$ is the integral which represents the open and closed strings channels.  In this amplitude $\veps_1$, $\veps_2$ and $\veps_3$ are the polarization of the RR, the gauge field, and the B-field, respectively. 

Using the totally antisymmetric property of the D$_p$-brane world volume form $\eps_{a_0\cdots a_p}$, one can easily rewrite the amplitude in terms of the B-field strength and the gauge field strength.
 Using the fact that the amplitude should satisfy the Ward identity corresponding to the NSNS and NS gauge symmetries, one realizes that the above coupling is the only possibility. So even without using the string theory calculation, one can find the above amplitude.  The string theory, however, gives information about the integral ${\cal Q}$ as well. The explicit form of this integral in terms of  the Mandelstam variables has been found  in \cite{Becker:2011ar}. The T-dual Ward identity then produces the following terms  \cite{Velni:2012sv}:
\beqa
A_2(C_{ij}^{(p-1)})&\sim&T_p(\veps_1 )_{ij}{}^{a_4\cdots a_p}\eps_{a_0\cdots a_p}p_3^ip_3^{a_3}[2(\veps_3^S)^{a_2j}p_2^{a_0}\veps_2^{a_1}+(\veps_3^A)^{a_1a_2}p_2^{a_0}\phi^j]{\cal Q}\nonumber\\
A_3(C_{ijk}^{(p+1)})&\sim&\frac{1}{2}T_p(\veps_1 )_{ijk}{}^{a_3\cdots a_p}\eps_{a_0\cdots a_p}p_3^ip_3^{a_2}[-2(\veps_3^A)^{jk}p_2^{a_0}\veps_2^{a_1}+4(\veps_3^S)^{a_1j}p_2^{a_0}\phi^k]{\cal Q}\nonumber\\
A_4(C_{ijkl}^{(p+3)})&\sim&T_p(\veps_1 )_{ijkl}{}^{a_2\cdots a_p}\eps_{a_0\cdots a_p}p_3^ip_3^{a_1}p_2^{a_0}(\veps_3^A)^{jk} \phi^l{\cal Q}\labell{A20}
\eeqa
On the other hand, the RR Ward identity connects the amplitude \reef{A11} to the second part in which the RR $(p-3)$-form has no transverse index.
 In this section we are going to find the T-dual multiplet corresponding to the second part. 

\subsection{RR $(p-3)$-form with no transverse index}
 
 The explicit calculation of the S-matrix element of the   RR $(p-3)$-form with no transverse index    gives the following result \cite{Becker:2011ar,Garousi:2012gh}:
\beqa
{A}_0&\!\!\!\!\!\sim\!\!\!\!\!&- (F_1)^{a_0 a_1}\bigg[ {\veps_2}^{a_3} p_3^{a_2} (p_1\inn D \inn \veps_3^A)^{a_4} J_1
+ {\veps_2}^{a_3} p_3^{a_2} (p_1\inn \veps_3^A)^{a_4}J_3+2 {\veps_2}^{a_3} p_3^{a_2} (p_2\inn \veps_3^A)^{a_4}J_2\nonumber\\
&&\qquad-\frac{1}{4}
 {\veps_2}^{a_2} p_3\inn V\inn p_3 (\veps_3^A)^{a_3 a_4} (J_1+J_3)+\frac{1}{2} p_3^{a_2} p_3\inn   \veps_2  (\veps_3^A)^{a_3 a_4} (J_1-2J_2+J_3)\bigg]\nonumber\\
   &&-(f_1)^{a_0i} {\veps_2}^{a_2} p_3^{a_1} {p_3}_i (\epsilon_3^A)^{a_3 a_4} \left(J_1-J_3\right)\labell{B0a1}
\eeqa
where the RR field strength $(F_1)^{a_0 a_1}=p_1^{a_0}{\veps_1}^{a_1}-p_1^{a_1}{\veps_1}^{a_0}$ and the RR factor $(f_1)^{a_0 i}=- p_1^{i}{\veps_1}^{a_0}$.  The diagonal matrix $D$ is $D=V-N$ where $V$ is the flat metric of the world volume space and $N$ is the flat metric of the transverse space. There is an overall factor of $T_4\eps_{a_0\cdots a_4}$. For simplicity we have written the amplitude for $p=4$. It can easily be extended to arbitrary $p$ by contracting  the extra word volume indices with the RR potential. The closed string and the open string channels appear in the integrals  $J_1, J_2, J_3$. The explicit form of these integrals have been found in \cite{Garousi:2012gh}. 

Note that $(f_1)^{a_0 i}$ in the last line of \reef{B0a1} is not the RR field strength. In fact the RR Ward identity connects the amplitude \reef{A11} to the last term in \reef{B0a1}, so there is the following relation between ${\cal Q}, J_1,\, J_3$:
\beqa
{\cal Q}&=&J_1-J_3
\eeqa
which can be verified using the explicit form of these integrals.  The last term in \reef{B0a1}, however, breaks the NS gauge symmetry. The RR gauge invariant terms in the first two lines are needed to make this term to be invariant under the NS and the  NSNS gauge transformations. These constraints give the following  relation between the integrals: 
\beqa
2p_1\inn N\inn p_3 (J_1-J_3)+p_3\inn V\inn p_3 (J_1+J_3)+2 p_2\inn p_3 (J_1-2J_2+J_3)&=&0\labell{id21} 
\eeqa
Therefore, there are two independent integrals in the amplitude \reef{B0a1}. 

One can verify that the terms in the first two lines of \reef{B0a} are all possible independent contractions  between  $(F_1)^{a_0a_1}$, the B-field, the gauge field,   two momenta and $\eps_{a_0a_1a_2a_3a_4}$. One may  consider the terms ${\veps_2}^{a_2} p_2\inn p_3 (\veps_3^A)^{a_3 a_4} $ or  ${\veps_2}^{a_2} p_1\inn N \inn p_3 (\veps_3^A)^{a_3 a_4} $ as well. However, before fixing the integrals, these terms can be absorbed into the fourth term in \reef{B0a1}. Therefore,   the string theory calculates the coefficients of all independent terms such that the amplitude satisfies  various Ward  identities. The coefficients  have  information about the open and closed string poles as well.     As we will see in the next subsections, the T-dual Ward identity which connects the above amplitude to all other RR potential,  does not produce any new integrals.

To apply the T-dual Ward identity on the amplitude \reef{B0a1}, it is convenient to rewrite the amplitude in terms of the flat metrics $V,\, N$. Using the relations $D=V-N$ and $\eta=V+N$, one finds
\beqa
{A}_0 &\!\!\!\!\!\sim\!\!\!\!\!& (F_1)^{a_0 a_1}\bigg[ {\veps_2}^{a_3} p_3^{a_2} (p_1\inn N \inn \veps_3^A)^{a_4} \cQ
+ {\veps_2}^{a_3} p_3^{a_2} (p_2\inn \veps_3^A)^{a_4}\cQ_2\nonumber\\
&& \qquad\qquad+{\veps_2}^{a_3} p_3^{a_2} (p_3\inn V\inn \veps_3^A)^{a_4}\cQ_1+\frac{1}{4}
 {\veps_2}^{a_2} p_3\inn V\inn p_3 (\veps_3^A)^{a_3 a_4} \cQ_1\nonumber\\
&& \qquad\qquad-\frac{1}{2} p_3^{a_2} p_3\inn   \veps_2  (\veps_3^A)^{a_3 a_4} \cQ_2\bigg]\nonumber\\
   &&-(f_1)^{a_0i} {\veps_2}^{a_2} p_3^{a_1} {p_3}_i (\epsilon_3^A)^{a_3 a_4} {\cal Q}\labell{B0a}
\eeqa
where 
\beqa
\cQ_1=J_1+J_3&;&\cQ_2=J_1-2J_2+J_3
\eeqa
The identity \reef{id21} then becomes
\beqa
2p_1\inn N\inn p_3 \cQ+p_3\inn V\inn p_3 \cQ_1+2 p_2\inn p_3 \cQ_2&=&0\labell{id2clos1open} 
\eeqa
One may write $\cQ=p_3\inn V\inn p_3\,p_2\inn p_3\cQ'$, $\cQ_1=p_1\inn N\inn p_3\,p_2\inn p_3\cQ_1'$ and $\cQ_2=p_1\inn N\inn p_3\,p_3\inn V\inn p_3\cQ_2'$. Then the above constraint can be solved to write the amplitude \reef{B0a} in terms of two integrals. However, we prefer to work with the three integrals and the constraint \reef{id2clos1open}.

\subsubsection{RR $(p-1)$-form with one transverse index}

In this section we are going to apply the T-dual Ward identity on the amplitude  \reef{B0a}. We have reviewed the method for applying the linear T-duality on the scattering amplitudes (the T-dual Ward identity)  in section 2. We refer the interested readers to \cite{Velni:2012sv} for more details on how to apply it to the specific cases.  Following \cite{Velni:2012sv}, one finds the amplitude \reef{B0a} is covariant under the linear T-duality when the isometric index  $y$ is carried by the RR potential. However, when the $y$-index is carried by the NS or the  NSNS polarizations, it is not invariant. Using the same steps as we have done in \cite{Velni:2012sv} , one finds that  the following amplitude has to be add to the  amplitude \reef{B0a} to make it invariant  under the linear T-duality transformations:
\beqa
{A}_1 &\sim&(f_1)^{a_0 a_1}{}_{ i} \bigg[-\frac{1}{2} (\veps_2)^{a_3} p_3\inn V\inn p_3 (\veps_3^S)^{a_2 i} \cQ_1- p_3^{a_2} p_3\inn V\inn \veps_2  (\veps_3^S)^{a_3 i}\cQ_2\nonumber\\
&&\qquad\qquad+\frac{1}{4}  \phi ^i p_3\inn V\inn p_3 (\veps_3^A)^{a_2 a_3}\cQ_1+ (\veps_2)^{a_3} p_3^{a_2} (p_1\inn N\inn \veps_3^S)_i\cQ\nonumber\\
&&\qquad\qquad-\phi ^i p_3^{a_2} (p_1\inn N\inn \veps_3^A)^{a_3}\cQ + {\veps_2}^{a_3} p_3^{a_2} (p_2\inn V\inn\veps
_3^S)_i\cQ_2\nonumber\\
&&\qquad\qquad+ {\veps_2}^{a_3} p_3^{a_2} (p_3\inn V\inn \veps_3^S)_i\cQ_1- \phi ^i p_3^{a_2} (p_2\inn V\inn \veps_3^A)^{a_3}\cQ_2\nonumber\\
&&\qquad\qquad- \phi ^i p_3^{a_2} (p_3\inn V\inn \veps_3^A)^{a_3}\cQ_1\bigg]\nonumber\\
&&- (f_1)^{a_0}{}_{ ij} p_3^{{a_1}} {p_3}^i\bigg[\phi ^ j (\veps_3^A)^{{a_2} {a_3}}-2 {\veps_2}^{a_3} (\veps_3^S)^{a_2 j}\bigg]\cQ\labell{B1a}
\eeqa
where the RR factors $(f_1)^{a_0 a_1 i}=-2 p_1^{a_1}{\veps_1}^{a_0 i}$ and $(f_1)^{a_0i j}=p_1^{j}{\veps_1}^{a_0 i}-p_1^{i}{\veps_1}^{a_0 j}$. For simplicity we have written the amplitude for $p=3$. In above amplitude $\phi$ is the polarization of the transverse scalar fields, and $\veps^S$ is the polarization of the graviton. Note that each term has either one transverse polarization which is the T-duality of the gauge field, or one symmetric NSNS polarization which is the T-dual of the antisymmetric NSNS polarization in \reef{B0a}. 
The above amplitude satisfies the Ward identity corresponding to the antisymmetric NSNS and  the NS  gauge symmetries. They are inherited from the amplitude \reef{B0a}. The graviton term in the last line also satisfies the Ward identity corresponding to the symmetric NSNS gauge transformation. However, the other graviton terms   do not satisfy this Ward identity.  

This indicates that the T-dual Ward identity could not capture all terms of the scattering amplitude of the RR $(p-1)$. In fact under the T-duality, the RR potential  $C^{(n)}$ which has no $y$ index, transforms   to  $(C^{(n+1)})^y$ which has one $y$-index. On the other hand, if this $y$-index is contracted with a polarization of the NSNS or the NS state in \reef{B0a}, the T-duality then can  capture it. However, if the $y$-index is contracted with the momentum of the NSNS polarization tensor in \reef{B0a}, then the T-duality can not capture it because in the T-duality transformation it is implicitly assumed that field are independent of the $y$-coordinate. Therefore, the T-dual Ward identity can not capture the terms which have the RR factor $(f_1^{(n+1)})^ip_{3i}$. The terms in the second bracket in  \reef{B1a} have already one $p_{3i}$ which contracted with the RR factor. So it is impossible to have another $p_{3j}$ to contract the  RR factor. However, the terms in the first bracket have no $p_{3i}$, so it is possible to include terms which have $(f_1)^{a_0a_1i}p_{3i}$. These terms could not be captured by the T-dual Ward identity.  

To find such terms, 
we can consider all   independent terms made of one momentum,  $\veps_2,\,\veps_3^S$ or $\phi_2,\,\veps_3^A$ which carry the indices $(\cdots)^{a_2a_3}$. Each term should be invariant under the linear T-duality when the world volume indices $a_2$ and $ a_3$ are not the $y$-index, \eg $(\veps_2\inn V \inn \veps_3^S)^{a_3}- (\phi_2\inn N \inn \veps_3^A)^{a_3}$ is invariant under the linear T-duality when $a_3\neq y$. Choosing all such terms which are 7 terms,  with unknown coefficients and imposing the condition that they should satisfy the Ward identity corresponding to the  NSNS and the NS gauge transformations, one finds the following result: 
\beqa
{A'}_1  &\!\!\!\!\!\sim\!\!\!\!\!&(f_1 )^{{a_0}{a_1}i} {p_3}_i \bigg[-\frac{1}{2} p_3^{{a_2}} {\veps_2}^{{a_3}} {\Tr}[\veps_3^S \inn V]\cQ_1 +p_3^{{a_2}}\cQ_2 \bigg((\veps_2 \inn V\inn \veps_3^S)^{{a_3}}-(\phi \inn N\inn \veps_3^A)^{{a_3}}\bigg)\nonumber\\
&&\qquad+\frac{1}{2} p_3\inn N\inn \phi  (\veps_3^A)^{{a_2}{a_3}}\cQ_2- {\veps_2}^{{a_3}} (p_1\inn N\inn \veps_3^S)^{{a_2}}\cQ - {\veps_2}^{{a_3}}(p_2\inn V\inn \veps_3^S)^{{a_2}}\cQ_2\bigg]\labell{B1add}
\eeqa
The combination of the above amplitude and amplitude \reef{B1a} satisfies the Ward identity corresponding to the NSNS and  the NS gauge transformation. They satisfy the T-dual Ward identity when the $y$-index is carried by RR potential. Otherwise they are not invariant under the linear T-duality. In the next subsection, we will find the amplitudes which are needed to make the amplitudes \reef{B1a} and \reef{B1add} to be invariant under the T-duality. 

\subsubsection{RR $(p+1)$-form with two transverse indices}

The amplitude \reef{B1a} makes the amplitude \reef{B0a} to be invariant under the linear T-duality when the $y$-index in the amplitude \reef{B0a} is carried by the NSNS and the NS polarization tensors. However, the amplitude \reef{B1a} is invariant under the T-duality only when the $y$-index is carried by the RR potential, otherwise it is not invariant. To fix this problem, we have to add the following amplitude to it: 
\beqa
{A}_2 &\sim&(f_1)^{{a_0}{a_1}}{}_{ij}\bigg[\frac{1}{4} p_3\inn V\inn p_3 \bigg(\veps_2^{{a_2}} (\veps_3^A)^{ij}+2 \phi ^i (\veps_3^S)^{{a_2}j}\bigg) \cQ_1\nonumber\\
&&\qquad\qquad-\frac{1}{2}p_3^{{a_2}} p_3\inn V\inn \veps_2  (\veps_3^A)^{ij}\cQ_2+ \phi ^j p_3^{{a_2}} (p_1\inn N\inn \veps_3^S)^i\cQ\nonumber\\
&&\qquad\qquad+ \phi ^j p_3^{{a_2}} (p_2\inn V\inn \veps_3^S)^i \cQ_2+ \phi ^j p_3^{{a_2}} (p_3\inn V\inn \veps_3^S)^i\cQ_1\bigg]\nonumber\\
&&-(f_1)^{{a_0}}{}_{ijk}  p_3^{{a_1}} {p_3}^i  \bigg[{\veps_2}^{{a_2}} (\veps_3^A)^{jk}+2 \phi ^j (\veps_3^S)^{{a_2}k}\bigg]\cQ\labell{B2a} 
\eeqa
where   $(f_1)^{a_0 a_1 ij}=p_1^{a_0}{\veps_1}^{a_1 ij}- p_1^{a_1}{\veps_1}^{a_0 ij}$ and $(f_1)^{a_0i j k}=-p_1^{k}{\veps_1}^{a_0 ij}-p_1^{j}{\veps_1}^{a_0 ki}-p_1^{i}{\veps_1}^{a_0 jk}$. For simplicity we have written the amplitude for $p=2$. Note that the RR factors are not the RR field strengths.
The above amplitude which has been found by imposing the T-dual Ward identity on the amplitude  \reef{B1a}, is not the full amplitude for the RR $(p+1)$-form with two transverse indices because it is not invariant under the NSNS and the NS gauge transformations. However, the terms in the last line satisfy these Ward identities so the T-dual Ward identity could captured all terms which have the RR factor $(f_1)^{a_0i j k}$. As in the previous section, there should be  some terms in the first bracket which are proportional to  $(f_1)^{{a_0}{a_1}ij}p_{3j}$. These term  are not captured by the T-duality.  

One may either impose the Ward identity corresponding to the NSNS and the NS gauge transformations to find the gauge completion of the   amplitude in the first bracket in \reef{B2a}, as we have done in the previous section. Or  one may impose the T-dual Ward identity to find  the T-dual completion of the amplitude \reef{B1add} when the $y$-index is carried by the NSNS and the NS polarization tensors. In both cases one finds the following result:
\beqa
{A'}_2 &\sim& ({f_1} )^{{a_0}{a_1}ij}{p_3}_i \bigg[-\frac{1}{2}(p_3)^{{a_2}} \phi _j {\Tr}[\veps_3^S\inn V] \cQ_1+  p_3\inn N\inn \phi  (\veps_3^S)^{{a_2}}{}_{j}\cQ_2\nonumber\\
&&\qquad\qquad\quad+ p_3^{{a_2}}\bigg((\veps_2\inn V\inn \veps_3^A)_j-(\phi \inn N\inn \veps_3^S)_j\bigg)\cQ_2\nonumber\\
 &&\qquad\qquad\quad-\bigg(\phi _j (p_1\inn N\inn \veps_3^S)^{{a_2}}-{\veps_2}^{{a_2}} (p_1\inn N\inn \veps_3^A)_j\bigg)\cQ\nonumber\\
 &&\qquad\qquad\quad-\bigg(\phi _j (p_2\inn V\inn \veps_3^S)^{a_2}-{\veps_2}^{{a_2}} (p_2\inn V\inn \veps_3^A)_j\bigg)\cQ_2\bigg]\labell{B2add} 
\eeqa
The combination of the above amplitude and the amplitude \reef{B2a} satisfies the Ward identities corresponding to the NSNS and the NS gauge transformations. Nigher the above  amplitude nor  the amplitude \reef{B2a} are  invariant under linear T-duality when the $y$-index is carried by the NSNS and the NS polarization tensors in these amplitudes. In the next subsection we will find the T-dual completion of these amplitudes.

 
 \subsubsection{RR $(p+3)$-form with three transverse indices}
 The symmetric NSNS   and the gauge field  polarization tensors in the first and last line of the amplitude \reef{B2a} carry the world volume index $a_2$. So when the world volume index   $y$ is carried by these tensors, the amplitude does not satisfy the T-dual Ward identity.   So we must add the following amplitude to make it consistent with T-dual Ward identity:
 \beqa
{A}_3 &\sim& (f_1)^{{a_0}{a_1}ijk} \Bigg[ \frac{1}{4}\phi _k p_3\inn V\inn p_3 (\veps_3^A)_{ij}\cQ_1\bigg]\nonumber\\
&&- (f_1)^{{a_0}ijkl} \bigg[ \phi _l p_3^{a_1} {p_3}_i (\veps_3^A)_{jk}\cQ\bigg]\labell{B3a}
\eeqa
where $(f_1)^{{a_0}{a_1}ijk}=-2 p_1^{a_1}{\veps_1}^{a_0 ijk}$ and $(f_1)^{{a_0}ijkl}=p_1^{l}{\veps_1}^{a_0 ijk}-p_1^{k}{\veps_1}^{a_0 ikl}+p_1^{j}{\veps_1}^{a_0ikl}-p_1^{i}{\veps_1}^{a_0jkl}$. For simplicity we have written the amplitude for $p=1$. Similarly, we have to add the following amplitude to \reef{B2add} to make it invariant under the linear T-duality:
\beqa
{A'}_3 &\sim& \frac{1}{2} ({f_1} )^{{a_0}{a_1}ijk}{p_3}_i \bigg[ p_3 \inn N\inn \phi  (\veps_3^A)_{jk}\cQ_2\nonumber\\
&&\qquad\qquad  \qquad-2 \phi _j (p_1\inn N\inn \veps_3^A)_k\cQ -2 \phi _j (p_2\inn V\inn \veps_3^A)_k\cQ_2\bigg]\labell{B3add} 
\eeqa
The combination of the above two amplitudes   satisfies the Ward identity corresponding to the antisymmetric NSNS gauge transformation.  
 The antisymmetric NSNS polarization tensor in these amplitudes does not  carry any world volume index. So the above amplitudes are invariant under linear T-duality. So there is no need for the amplitude $A_4(C_{ijkl}^{(p+5)})$ to be added. In fact, noting that the open string momentum must be along the world volume directions, one can verify that it is impossible to have contraction between $\eps_{a_0\cdots a_p}$, one $C_{ijkl}^{(p+5)}$, three momenta, one NSNS and one NS polarization tensors. 

Therefore the amplitudes that we have found so far, \ie
\beqa
\matrix{A_0(C ^{(p-3)})& \!\!\!\!\!\rightarrow \!\!\!\!\!& A_1(C_{ i}^{(p-1)})& \!\!\!\!\!\rightarrow \!\!\!\!\!&A_2(C_{ ij}^{(p+1)})&  \!\!\!\!\!\rightarrow \!\!\!\!\! & A_3(C_{ ijk}^{(p+3)})\cr 
&&\downarrow & &\downarrow&& \downarrow\cr
&&  {A_1'} (C_{ i}^{(p-1)})& \!\!\!\!\!\rightarrow \!\!\!\!\!& {A_2'} (C_{ ij}^{(p+1)})& \!\!\!\!\!\rightarrow \!\!\!\!\!& {A_3'} (C_{ ijk}^{(p+3)})\cr} \labell{2close1open2} 
\eeqa
satisfy the Ward identity corresponding to the T-duality, the NSNS and the NS gauge transformations. However, they do not satisfy the Ward identity corresponding to the RR gauge transformation. In the next section we find some other  amplitudes by imposing the constraint that the amplitudes must satisfy the RR Ward identity as well. 

 \subsection{RR Ward identity}

In this section we are going to add the appropriate amplitudes to the amplitudes that have been found in the previous section to  make them satisfy the Ward identity corresponding to the RR gauge transformations as well as the T-duality and the NSNS and the NS gauge transformations. The RR Ward identity allows us to write the amplitudes in terms of the RR field strengths. 
The combination of the amplitudes $A_1(C_i^{(p-3)})$ in \reef{B0a} and $A_0(C^{(p-3)})$ in \reef{A11}, satisfies the RR Ward identity because they are the amplitudes which have been calculated explicitly in string theory \cite{Becker:2011ar,Garousi:2012gh}. The terms in the first three lines of \reef{B0a} are in terms of the RR field strength $F_1^{a_0a_1}$. The combination of the terms in the last line of \reef{B0a} and \reef{A11} can also be written in terms of the RR field strength $F_1^{a_0i}=p_1^{a_0}\veps_1^i-p_1^i\veps_1^{a_0}$.

Using the T-dual Ward identity, we have found the amplitudes for the RR potential $(p-1)$-form which has two and one transverse indices. One can verify that it is impossible to have the amplitude for the RR $(p-1)$-form which carries more that two transverse indices. However, there are possibilities for having amplitude for the RR $(p-1)$-form which carries zero transverse index. This amplitude can be found by imposing the RR Ward identity on the amplitudes that we have found in the previous section. The combination of the terms in the last line of \reef{B1a} and the terms in the first line of \reef{A20} satisfies the RR Ward identity, \ie they can be written as in the last line of \reef{B1a} but with the RR field strength $(F_1)^{a_0ij}=p_1^{a_0}\veps_1^{ij}+p_1^j\veps_1^{a_0i}+p_1^i\veps^{j a_0}$ instead of $(f_1)^{a_0ij}$. 

However, the other terms in \reef{B1a} and the terms in \reef{B1add} do not not satisfy the RR Ward identity because the RR factor $(f_1 )^{{a_0}{a_1}i}$ is not the full RR field strength. So the obvious extension of these amplitudes to the RR invariant amplitudes is  to extend this factor to the RR field strength $(F_1 )^{{a_0}{a_1}i}=p_1^{a_0}\veps_1^{a_1 i}-p_1^{a_1}\veps_1^{a_0 i}+p_1^i \veps_{1}^{a_0a_1}$.   However, the new amplitude resulting from the last term in $(F_1 )^{{a_0}{a_1}i}$ would not be invariant under the NSNS and the NS gauge transformations. To remedy this failure, one has to still add another amplitude which should be proportional to the RR field strength $ (F_1)^{a_0a_1a_2}$ and should make the above terms to be invariant under the NSNS and the NS gauge transformations. We consider all independent terms $(\cdots)^{a_3}$ containing two momenta and the NSNS and the NS polarization tensors which are invariant under the linear T-duality when the world volume index $a_3$ is not the $y$-index. Fixing the coefficients of these terms  by combining them with the above non-gauge invariant terms and requiring that they should satisfy the Ward identity corresponding to the NSNS and the NS gauge transformations, one finds the following result: 
\beqa
{A}_0 &\!\!\!\!\!\sim\!\!\!\!\!&  (F_1 )^{{a_0} {a_1} {a_2}}\bigg[ \frac{1}{3}\veps_2^ {{a_ 3}}\bigg(3p_ 2\inn V\inn \veps_3^S \inn V\inn p_2 \cQ_ 3 +p_1\inn N\inn \veps_3^S \inn N\inn p_1  \cQ + p_1\inn N\inn \veps_3^S \inn V\inn p_3\cQ_1\nonumber\\
&& +3p_ 2\inn V\inn \veps_3^S \inn V\inn p_3 \cQ_ 4 + 2p_1\inn N\inn \veps_3^S \inn V\inn p_2 \cQ_2 -\frac{1}{2} (p_ 1\inn N\inn p_3 \cQ_ {1}+3p_2\inn p_3 \cQ_4){\Tr}[{\veps_3^S}\inn V]\bigg)\nonumber\\
&& -\frac {1}{3} p_3\inn V\inn {\veps_2}\bigg((p_ 1\inn N\inn \veps_3^S)^ {{a_ 3}} \cQ_2+3 (p_ 2\inn V\inn \veps_3^S)^ {{a_ 3}}\cQ_3+\frac{3}{2}p_3^{a_3}{\Tr}[{\veps_3^S}\inn V]\cQ_4\bigg)\labell{Fp}\\
&& -\frac {1}{3} p_3\inn N\inn \phi\bigg( (p_ 1\inn N\inn \veps_3^A)^{a_ 3}\cQ_2+3(p_ 2\inn V\inn \veps_3^A)^{{a_ 3}}\cQ_3+3(p_ 3\inn V\inn \veps_3^A)^{{a_ 3}}\cQ_4\bigg)\nonumber\\
&& -\frac{1}{2}p_3\inn V\inn p_3\bigg((\veps_2 \inn V\inn \veps_3^S)^{a_3} - (\phi \inn N\inn \veps_3^A)^{a_3} \bigg)\cQ_4 + p_ 3^{a_ 3}\bigg((p_2\inn V\inn \veps_3^S \inn V\inn \veps_2 +p_ 2\inn V\inn \veps_3^A \inn N\inn \phi) \cQ_3\nonumber\\
&&  +(p_3\inn V\inn \veps_3^S \inn V\inn \veps_2 +p_ 3\inn V\inn \veps_3^A \inn N\inn \phi) \cQ_4 + \frac {1} {3}  (p_ 1\inn N\inn \veps_3^S \inn V\inn \veps_2+p_1\inn N\inn \veps_3^A \inn N\inn \phi ) \cQ_2\bigg)\bigg]\nonumber
\eeqa
where the new integrals $\cQ_3,\,   \cQ_4$   satisfy the following   relation:
\beqa
3p_3\inn V\inn p_3 \cQ_4+6p_2\inn p_3\cQ_3+2p_1\inn N\inn p_3\cQ_2&=&0\labell{2cons}
\eeqa
The above constraint can not be used to find the integrals  $\cQ_3,\,   \cQ_4$. Unlike the  Ward identities corresponding to the T-duality and the NSNS/NS gauge transformations which do not produce new integrals, the Ward identities corresponding to the RR and the NSNS/NS gauge transformations produce new integrals $\cQ_3,\,   \cQ_4$. In order to study the above amplitude at low energy, one has to perform the explicit string theory calculation to find these integrals.  In this paper, we are not interested in the explicit form of these integrals.

The  amplitude \reef{Fp} satisfies the T-dual Ward identity when the $y$-index is carried by the RR field strength, however, when $a_3$ is the $y$-index it is not invariant under the linear T-duality. To make \reef{Fp} invariant under the linear T-duality, one has to include the following amplitude: 
\beqa
{A}_1 &\!\!\!\!\!\sim\!\!\!\!\!&  (f_1 )^{{a_0} {a_1} {a_2}}{}_{i}\bigg[ \frac{1}{3}\phi^ {i}\bigg(3p_ 2\inn V\inn \veps_3^S \inn V\inn p_2 \cQ_ 3 +p_1\inn N\inn \veps_3^S \inn N\inn p_1  \cQ + p_1\inn N\inn \veps_3^S \inn V\inn p_3 \cQ_1\nonumber\\
&& + 2p_1\inn N\inn \veps_3^S \inn V\inn p_2 \cQ_2 + 3p_2\inn V\inn \veps_3^S \inn V\inn p_3 \cQ_4 -\frac{1}{2} (p_ 1\inn N\inn p_3 \cQ_ {1}+3p_2\inn p_3 \cQ_4){\Tr}[{\veps_3^S}\inn V]\bigg)\nonumber\\
&& -\frac {1}{3} p_3\inn V\inn {\veps_2}\bigg((p_ 1\inn N\inn \veps_3^A)^ {i} \cQ_2+3 (p_ 2\inn V\inn \veps_3^A)^ {i}\cQ_3\bigg)-\frac{1}{2}p_3\inn V\inn p_3\bigg((\veps_2 \inn V\inn \veps_3^A)^{i} - (\phi \inn N\inn \veps_3^S)^{i} \bigg)\cQ_4\nonumber\\
&& -\frac{1}{3} p_3\inn N\inn \phi \bigg((p_ 1\inn N\inn \veps_3^S)^{i} \cQ_2+3(p_ 3\inn V\inn \veps_3^S)^{i} \cQ_4+3(p_ 2\inn V\inn \veps_3^S)^{i})\cQ_3\bigg) \bigg]\labell{Fp2}
\eeqa
where $ (f_1 )^{{a_0} {a_1} {a_2}i}=p_1^{a_0}\veps_1^{a_1a_2i}+p_1^{a_2}\veps_1^{a_0a_1i}+p_1^{a_1}\veps_1^{a_2a_0i}$. The  world volume form does not contracted with the NSNS or the NS polarization tensors, so the above  amplitude satisfies the T-dual Ward identity. However, it does not satisfy the Ward identity corresponding to the NSNS or the NS gauge transformations. So there are some missing terms which are not captured by the T-dual Ward identity. The missing terms are the following:
\beqa
{A'}_1 &\!\!\!\!\!\sim\!\!\!\!\!&  (f_1 )^{{a_0} {a_1} {a_2}}{}^{i}p_{3i}\bigg[\frac{1}{3}\bigg(p_1 \inn N\inn \veps_3^A\inn V\inn\veps_2 +p_1\inn N\inn\veps_3^S\inn N\inn \phi   \bigg)\cQ_2+\frac{1}{2}p_3\inn N \inn\phi {\Tr}[{\veps_3^S}\inn V]\cQ_4\nonumber\\
&& \qquad\qquad\qquad+\bigg(p_2 \inn V\inn \veps_3^A\inn V\inn\veps_2 +p_2\inn V\inn\veps_3^S\inn N\inn \phi   \bigg)\cQ_3\bigg]\labell{missing}
\eeqa
The combination of the above amplitude and \reef{Fp2} is invariant under the NSNS and the NS gauge transformations.

Since the RR factor  $ (f_1 )^{{a_0} {a_1} {a_2}i}$ is not the RR field strength, the   amplitudes $A_1+A_1'$ does not satisfy the RR Ward identity. It  can be easily extended to satisfy this Ward identity by extending the RR factor to the RR field strength  $ (F_1 )^{{a_0} {a_1} {a_2}i}$. The new terms in this amplitude, \ie the terms proportional to $(F_1 )^{{a_0} {a_1} {a_2}i}-(f_1 )^{{a_0} {a_1} {a_2}i}$, should satisfy the Ward identity corresponding to the NSNS and the NS gauge transformations.  

We now study the RR Ward identity of  the amplitudes in section 3.1.2. The combination of the terms in the last line of \reef{B2a} and in the second line of \reef{A20} satisfies the RR Ward identity. They can be written as the last line in \reef{B2a} in which the RR factor is replaced by the RR field strength $(F_1)^{a_0ijk}$. The terms in the first three lines of \reef{B2a} and the terms in \reef{B2add}, however,  do not satisfy the RR Ward identity because the RR factor in these amplitudes, \ie $ (f_1 )^{{a_0} {a_1}  ij}$,  is not the RR field strength.  They can easily be extended to the RR invariant amplitudes by extending the RR factor to the RR field strength $ (F_1 )^{{a_0} {a_1}  ij}$. The new terms in this amplitude, \ie the terms proportional to $(F_1 )^{{a_0} {a_1}  ij}-(f_1 )^{{a_0} {a_1}  ij}$, should satisfy the Ward identity corresponding to the NSNS and the NS gauge transformations. In the next section, we will show when the NSNS state is antisymmetric, the amplitude at order $\alpha'^2$ which has only contact terms,  can be written in terms of   field strengths.

The RR Ward identity of  the amplitudes in section 3.1.3 is as follows: The combination of terms in the last line of \reef{B3a} and in the third line of \reef{A20} satisfies the RR Ward identity. They can be written as the last line in \reef{B3a} in which the RR factor is replaced by the RR field strength $(F_1)^{a_0ijkl}$. The terms in the first   line of \reef{B3a} and the terms in \reef{B3add}, however,  do not satisfy the RR Ward identity because the RR factor in these amplitudes, \ie  $ (f_1 )^{{a_0} {a_1}  ijk}$,  is not the RR field strength.  They can easily be extended to the RR invariant amplitudes by extending the RR factor to the RR field strength $ (F_1 )^{{a_0} {a_1}  ijk}$. The new terms, \ie the terms proportional to  $(F_1 )^{{a_0} {a_1}  ijk}-(f_1 )^{{a_0} {a_1}  ijk}$  should satisfy the Ward identity corresponding to the NSNS and the NS gauge transformations.

Therefore the amplitudes that are invariant under the linear T-duality and all the gauge transformations can be written as three multiplets in terms of the RR field strength. The first multiplet is the following:
\beqa
 {A}_1(F_i^{(p-2)})\rightarrow {A}_2(F_{ij}^{(p)})\rightarrow {A}_3(F_{ijk}^{(p+2)})\rightarrow {A}_4(F_{ijkl}^{(p+4)})\labell{Fi}
 \eeqa
where $A_1,\, A_2,\,A_3$ and $A_4$ are the terms in the last lines of \reef{B0a}, \reef{B1a}, \reef{B2a} and \reef{B3a}, respectively, in which the RR factor $f_1$ is replaced by the RR field strength $F_1$. The other multiplet is 
 \beqa
\matrix{A_0(F ^{(p-2)})& \!\!\!\!\!\rightarrow \!\!\!\!\!& A_1(F_{ i}^{(p)})& \!\!\!\!\!\rightarrow \!\!\!\!\!&A_2(F_{ ij}^{(p+2)})&  \!\!\!\!\!\rightarrow \!\!\!\!\! & A_3(F_{ ijk}^{(p+4)})\cr 
&&\downarrow & &\downarrow&& \downarrow\cr
&&  {A_1'} (F_{ i}^{(p)})& \!\!\!\!\!\rightarrow \!\!\!\!\!& {A_2'} (F_{ ij}^{(p+2)})& \!\!\!\!\!\rightarrow \!\!\!\!\!& {A_3'} (F_{ ijk}^{(p+4)})\cr} \labell{Fii} 
\eeqa
where $A_0$ appears in the first bracket in \reef{B0a}. The amplitudes $A_1$, $A_2$ and $A_3$ appear in the first brackets in \reef{B1a}, \reef{B2a} and \reef{B3a}, respectively, in which the RR factor $f_1$ is replaced by the RR field strength $F_1$. The amplitudes $A_1'$, $A_2'$ and $A_3'$ are the amplitudes in \reef{B1add}, \reef{B2add} and \reef{B3add},  respectively, in which the RR factor $f_1$ is replaced by the RR field strength $F_1$. The last multiplet is 
\beqa
\matrix{A_0(F ^{(p)})& \!\!\!\!\!\rightarrow \!\!\!\!\!& A_1(F_{ i}^{(p+2)})  \cr
&& \downarrow\cr
&&  {A_1'} (F_{ i}^{(p+2)})\cr
} \labell{Fiii} 
\eeqa
where $A_0$ appears in \reef{Fp}. The amplitudes $A_1$ and $A_1'$ are the same as the amplitudes \reef{Fp2} and \reef{missing}, respectively,  in which the RR factor $f_1$ is replaced by the RR field strength $F_1$.

\subsection{Low energy couplings}

The S-matrix elements that we have found in the previous sections, can be analyzed  at low energy to extract the appropriate couplings in field theory at order $\alpha'^2$. To this end, we need the $\alpha'$-expansion of  the integrals that appear in the amplitudes. The  $\alpha'$-expansion of the integrals $J_1,\, J_2,\, J_3$ have been found in \cite{ Garousi:2012gh}. Using the relation $p_1\inn D\inn p_1+4p_1\inn p_2=p_3\inn D\inn p_3$, one finds
\beqa
J_1&=&-\frac{1}{p_1.p_3}-\frac{4}{p_3.D.p_3}+\frac{2}{3} \pi ^2 p_1.p_3+\frac{1}{6} \pi ^2 p_3.D.p_3 -\frac{8 \pi ^2 \left(p_2.p_3\right){}^2}{3 p_3.D.p_3} +\cdots\nonumber\\
 J_2&=&-\frac{1}{p_1.p_3}-\frac{4}{p_3.D.p_3}+\frac{2}{3} \pi ^2 p_1.p_3-\frac{1}{6} \pi ^2 p_1.D.p_1+\frac{1}{3} \pi ^2 p_3.D.p_3 -\frac{8 \pi ^2 \left(p_2.p_3\right){}^2}{3 p_3.D.p_3} +\cdots\nonumber\\
 J_3&=&-\frac{3}{p_1.p_3}-\frac{4}{p_3.D.p_3}+\frac{2}{3} \pi ^2 p_1.p_3+\frac{1}{2} \pi ^2 p_3.D.p_3 -\frac{8 \pi ^2 \left(p_2.p_3\right){}^2}{3 p_3.D.p_3} +\cdots\labell{expand}
\eeqa
where dots refer to the terms with more than two momenta. They are related to the couplings at order $O(\alpha'^3)$ in which we are not interested. 

It is interesting to note that the massless closed string pole $1/p_1\inn p_3$ appears only at the leading order which is resulted from the fact that there is neither the higher-derivative   couplings between three closed strings in the bulk nor the higher-derivative couplings between one closed and one open strings on the D-brane world volume.  The above expansions  can be used to find the low energy expansion of the integrals $\cQ,\, \cQ_1,\,\cQ_2$ which appear in the S-matrix elements in multiplets \reef{Fi} and \reef{Fii}. The massless poles at the leading order should be reproduced by the supergravity  couplings in the bulk, and the DBI and CS couplings on the D-brane. The next to the leading order terms  have massless open string pole and contact terms. The contact terms  do  not produce, in general,  the couplings of field theory at order $\alpha'^2$. One has to first calculate the massless pole in field theory which is produced by the  couplings of one closed and two open string states at order $\alpha'^2$ \cite{Hashimoto:1996kf,Garousi:1998fg}, and by the couplings of one closed and one open string states which are given by the DBI and CS actions. Then one should subtract it from the above massless pole. This subtraction may cancel some of the  contact terms as well.  The left over contact terms, then produces new couplings at order $\alpha'^2$ between one RR, one NSNS and one NS states in the field theory.

We are interested in this paper in finding the couplings of one  $F^{(p+2)}$ with two transverse indices, one B-field and one gauge boson. There is no coupling between one RR $(p+1)$-form and two gauge bosons \cite{ Garousi:1998fg}, so we expect the amplitude in the string theory side has no massless open string pole. The string theory amplitude is given by the sum of \reef{B2a} and \reef{B2add} in which the RR factors are replaced by the RR field strength and the NSNS polarization is antisymmetric. Using the expansion \reef{expand}, one can easily verify that it has no massless open string pole, as expected. The amplitude at order $\alpha'^2$ then has only contact terms. These contact terms   are the following:
\beqa
A_c(\alpha'^2)&\sim&\frac{ \pi ^2}{3}(F_1)^{{a_0}{a_1}}{}_{ij}\bigg[4 p_2.p_3 p_3^ip_3^{a_2} ({\veps_2}.V.{\veps_3^A})^j +2 p_3.V.p_3 p_3^ip_3^{a_2} ({\veps_2}.V.{\veps_3^A})^j  \nonumber\\
&&\qquad\qquad\quad-2 p_3.V.p_3 {\veps_2}^{{a_2}} p_3^i\left(p_1.N.{\veps_3^A}\right)^j +4 p_2.p_3 {\veps_2}^{{a_2}} p_3^i\left(p_2.V.{\veps_3^A}\right)^j  \nonumber\\
&&\qquad\qquad\quad+2 p_3.V.p_3 {\veps_2}^{{a_2}} p_3^i\left(p_2.V.{\veps_3^A}\right)^j -  p_2.p_3 p_3.V.p_3 {\veps_2}^{{a_2}} ({\veps_3^A})^{ij}  \nonumber\\
&&\qquad\qquad\quad+  p_1.N.p_3 p_3.V.p_3 {\veps_2}^{{a_2}} ({\veps_3^A})^{ij}  -2 p_2.p_3 p_3.V.{\veps_2} p_3^{a_2} ({\veps_3^A})^{ij}  \nonumber\\
&&\qquad\qquad\quad-  p_3.V.{\veps_2} p_3.V.p_3 p_3^{a_2} ({\veps_3^A})^{ij}  - 2 (p_2.p_3)^2   {\veps_2}^{{a_2}} ({\veps_3^A})^{ij}\bigg]\nonumber\\
&& +\frac{2 \pi ^2}{3}(F_1)^{{a_0}}{}_{ijk}  p_3.V.p_3 {\veps_2}^{{a_2}} p_3^ip_3^{a_1} ({\veps_3^A})^{jk} \labell{amp}
\eeqa
They satisfy the Ward identity corresponding to the gauge transformations.

The above amplitude is in terms of the RR field strength. Since they are  contact terms, one should be able to rewrite them in terms  of the field strengths $H=dB$ and $\cF=dA$ as well. To this end, we first write $p_3^{a_1}$ in the last line in terms of $-p_2^{a_1}-p_1^{a_1}$. The term in the last line corresponding to $-p_2^{a_1}$ can easily be written in terms of the field strengths. The term in the last line corresponding to  $-p_1^{a_1}$ can be combined with the terms in the first bracket to write them in terms of the field strengths. After some algebra one can write the above amplitude as
\beqa
A_c(\alpha'^2)&\sim& \frac{ \pi ^2 }{3}(F_1)^{{a_0}{a_1}}{}_{ij} p_1.V.p_1 (\cF_2.V.H_3)^{{a_2}ij}  -\frac{ \pi ^2}{9} (F_1)^{{a_0}}{}_{ijk} p_3.V.p_3 \cF_2^{{a_1}{a_2}} H_3^{ijk}  
\eeqa
It is interesting that the complicated amplitude \reef{amp} in terms of the polarization tensors has such a simple form in terms of the corresponding field strengths.

Transforming the above contact terms in the momentum space to the coordinate space, one finds the   following $\alpha'^2$ couplings on the world volume of the  D$_p$-brane:
\beqa
S&\!\!\!\!\!\supset\!\!\!\!&\pi^2\alpha'^2T_p\int d^{p+1}x\,\eps^{a_0\cdots a_p}\left( 
 \frac{1}{2!3!(p-1)!}{ F}^{(p+2)}_{a_2\cdots a_pikj} H^{ijk,a}{}_a( 2\pi\alpha'\cF_{a_0a_1}) \right.\nonumber\\
&&\qquad\qquad\qquad\qquad\qquad\left.+\frac{1}{2!p!}{ F}^{(p+2)}_{a_1\cdots a_pij,a}{}^a H^{bij}( 2\pi\alpha'\cF_{a_0b}) \right)\labell{first11}
\eeqa
where    the RR field strength is $F^{(n)}=dC^{(n-1)}$. The first term has been found in \cite{Velni:2012sv} by analyzing the amplitude for the RR $(p+1)$-form with three transverse indices.   The other coupling is a new coupling which should be added to the D$_p$-brane action at order $\alpha'^2$. 

Extending the above couplings to be covariant under the coordinate transformations and invariant under the   RR and NSNS gauge transformations, one finds the following nonlinear couplings at order $\alpha'^2$:
\beqa
S&\!\!\!\!\!\supset\!\!\!\!&\pi^2\alpha'^2T_p\int d^{p+1}x\,\eps^{a_0\cdots a_p}\left( 
 \frac{1}{2!3!(p-1)!}{\tilde{F}}^{(p+2)}_{a_2\cdots a_pikj} H^{ijk;a}{}_a{\tilde{B}}_{a_0a_1} \right.\nonumber\\
&&\qquad\qquad\qquad\qquad\qquad\left.+\frac{1}{4}\frac{1}{2!p!}{\tilde{F}}^{(p+2)}_{a_1\cdots a_pij;a}{}^a H^{bij}{\tilde{B}}_{a_0b} \right)\labell{first112}
\eeqa
where the nonlinear RR field strength and the NSNS gauge invariant ${\tilde{B}} $ are  
\beqa
\tilde{F}^{(n)}=dC^{(n-1)}+H\wedge C^{(n-3)}&;& {\tilde{B}}=B+ 2\pi\alpha'\cF\labell{tF}
\eeqa
The action  \reef{first112} predicts, among other things, the couplings with structure $C^{(p+1)}HB$. These couplings can be confirmed by the S-matrix elements of one RR and two NSNS states which we will find them in the next section.

There should be another term in the action \reef{first112} in which the RR field strength is ${\tilde{F}}^{(p+2)}_{a_0\cdots a_pi}$. This term is resulted from the low energy expansion of the amplitudes \reef{Fp2} and \reef{missing}. The explicit form of the integrals $\cQ_3,\, \cQ_4$ which appear in these amplitude,  can be found by the string theory calculation of the amplitudes \reef{Fp2} and \reef{missing} in which we are not interested in this paper.

 \section{ Three closed string amplitudes}

The disk-level S-matrix element of one RR $(p-3)$-form and two NSNS states has been calculated in \cite{Craps:1998fn,Garousi:2010bm,Garousi:2011ut,Becker:2011bw,Becker:2011ar}. The amplitude is nonzero when the RR polarization has two, one and zero transverse indices. The T-dual multiplets corresponding to the first two cases have been found in \cite{Velni:2012sv}. They satisfy the Ward identity corresponding to the T-dulity and the NSNS gauge transformations. However, since the RR $(p-3)$-form with no transverse index was not considered in this study, the multiplets found in \cite{Velni:2012sv} are not invariant under  the RR gauge transformations.  In this section we are going to consider the RR $(p-3)$-form with no transverse index. The requirement that this amplitude should satisfy the Ward identity corresponding to the T-duality and the NSNS gauge transformation allow us to find various S-matrix elements of one RR and two NSNS states.  

\subsection{RR $(p-3)$-form with no transverse index}
 
When the RR   $(p-3)$-form has no transverse index, the amplitude  is nonzero   for the case that the NSNS polarizations are both antisymmetric or symmetric \cite{Craps:1998fn,Garousi:2011ut,Becker:2011ar}. The amplitude can be written as two parts. One part is in terms of the RR field strength $(F_1)^{a_0a_1}=p_1^{a_0}\veps_1^{a_1}-p_1^{a_1}\veps_1^{a_0}$ and the other part is in terms of the RR factor $(f_1)^{ia_0}=p_1^i\veps_1^{a_0}$, \ie
\beqa
{\bf A}_0 &=& {  A}_0(F_1)+{\cal A}_0(f_1) \labell{A0}
\eeqa
The subscribe 0 refers to the number of transverse indices of the RR potential. For simplicity we  consider the case that $p=4$.

The second part in \reef{A0} which is non-zero when both NSNS states are antisymmetric, includes the following terms \cite{Garousi:2011ut} (see eq.(34) in \cite{Garousi:2011ut}\footnote{Note that there is a type in the last line of eq.(34): the overall factor 2 should be $1/4$.}):
\beqa
{\cal A}_0 (f_1)&\!\!\!\!\!\!\!\sim\!\!\!\!\!\!\!&\frac{1}{2}(f_1 )_i^{a_4 }\bigg[\frac{1}{2}p_3\inn V\inn\veps_2^{a_2}\veps_3^{a_1a_3}(p_2^ip_2^{a_0}\cI_3+p_3^ip_3^{a_0}\cI_2)+\frac{1}{2}p_3\inn N\inn\veps_2^{a_2}\veps_3^{a_1a_3}(p_3^ip_2^{a_0}\cI_3+p_2^ip_3^{a_0}\cI_2)\nonumber\\
&&\qquad\quad-2p_3^ip_3^{a_0}p_2\inn V\inn\veps_2^{a_2}\veps_3^{a_1a_3}\cI_7+p_3^ip_3^{a_0}p_1\inn N\inn\veps_2^{a_2}\veps_3^{a_1a_3}\cI_1 +(2\leftrightarrow 3)\nonumber\\
&&\qquad\quad+\frac{1}{4}\veps_2^{a_0a_1}\veps_3^{a_2a_3} \bigg(p_2^ip_3\inn V\inn p_3\cI_4+\frac{1}{2}p_3^ip_2\inn V\inn p_3\cI_2-\frac{1}{2}p_3^ip_2\inn N\inn p_3\cI_3\bigg)\bigg] \labell{A''0A} 
\eeqa
Our notation for the first $(2\leftrightarrow 3)$ in above amplitude and in all other amplitudes in this paper, is that the expressions from the beginning  up to that point, including the overall factor,  should be interchanged under $2\leftrightarrow 3$. In above equation,   $\cI_1,\,\cI_2,\,\cI_3,\, \cI_4$ and $\cI_7$ are the integrals that represent the open and closed string channels. The explicit form of these integrals are given in \cite{Garousi:2010bm,Garousi:2011ut}. They satisfy the  relation \cite{Garousi:2010bm}:
\beqa
&&-2p_1\inn N\inn p_2\cI_1+2p_2\inn V\inn p_2\cI_7+p_2\inn N\inn p_3\cI_3-p_2\inn V\inn p_3\cI_2=0\labell{id2}
\eeqa
 and similar relation under the interchange of $(2\leftrightarrow 3)$. The symmetries of the integrals under the interchange of $(2\leftrightarrow 3)$ are such that $\cI$ is invariant, $\cI_2\leftrightarrow\cI_3$ and $\cI_4\leftrightarrow\cI_7$. Using the above relations one finds that the amplitude \reef{A''0A} satisfies the Ward identity corresponding to the antisymmetric NSNS gauge transformation \cite{Garousi:2010bm}.  If one adds to \reef{A''0A}  the amplitude of the RR $(p-3)$-form with one transverse index, then the RR factor in the amplitude \reef{A''0A} is extended to the RR field strength $(F_1 )_i^{a_4 }$, \ie the amplitude ${\cal A}_0 (C)$ is extended to ${\cal A}_1 (F_i)$ where the subscript 1 in the latter amplitude refers to the number of the transverse index of the RR field strength. As a result, it would satisfy the RR Ward identity. The amplitude of the RR $(p-3)$-form with one transverse index has also some terms which are combined with the amplitude of the RR $(p-3)$-form with two transverse indices to become RR invariant \cite{Velni:2012sv}. The amplitudes in terms of the RR field strength, however, do not satisfy the NSNS Ward identity unless one consider the first part in \reef{A0}.

The first part of \reef{A0} which is non-zero when the NSNS states are both   symmetric or antisymmetric, is given as \cite{Garousi:2011ut}
\beqa
{  A}_0 &\!\!\!\!\!\sim\!\!\!\!\!& - \frac{1}{4}(F_1 )^{a_0a_4 }\bigg[2p_3^{a_1}\bigg(-p_2\inn N\inn p_3(\veps_2^S\inn N\inn\veps_3^S)^{a_2a_3}+p_2\inn V\inn p_3(\veps_2^S\inn V\inn\veps_3^S)^{a_2a_3}\nonumber\\
&&+(p_3\inn N\inn\veps_2^S)^{a_2}\,(p_2\inn N\inn\veps_3^S)^{a_3}-(p_3\inn V\inn\veps_2^S)^{a_2}\, (p_2\inn V\inn\veps_3^S)^{a_3}\bigg)\cJ-2p_2\inn V\inn\veps_2^{a_2}\veps_3^{a_1a_3}p_3\inn V\inn p_3\cJ_3\nonumber\\
&&+
p_3^{a_1}\bigg(
p_2\inn N\inn p_3(\veps_2\inn V\inn\veps_3)^{a_2a_3}-p_2\inn V\inn p_3(\veps_2\inn N\inn\veps_3)^{a_2a_3}-\veps_3^{a_2a_3}p_3\inn V\inn\veps_2\inn N\inn p_3\bigg)\cJ\nonumber\\
&&+\bigg(p_3\inn V\inn\veps_2\inn V\inn p_2\cJ_1-\frac{1}{2}p_3\inn V\inn\veps_2\inn N\inn p_1\cI_3
+p_2\inn V\inn\veps_2\inn N\inn p_3\cJ_2-p_3\inn V\inn\veps_2\inn N\inn p_3\cJ_5\nonumber\\
&&+\frac{1}{2}p_3\inn N\inn\veps_2\inn N\inn p_1\cI_2\bigg)p_3^{a_1}\veps_3^{a_2a_3}-p_3^{a_1}p_1\inn N\inn\veps_2^{a_2}\,p_2\inn V\inn\veps_3^{a_3}\cI_3-p_3^{a_1}p_1\inn N\inn\veps_2^{a_2}\,p_1\inn N\inn\veps_3^{a_3}\cI_1\nonumber\\
&&+p_3^{a_1}p_1\inn N\inn\veps_2^{a_2}\,p_2\inn N\inn\veps_3^{a_3}\cI_2+2p_3^{a_1}p_3\inn V\inn\veps_2^{a_2}\,p_2\inn N\inn\veps_3^{a_3}\cJ_5+4p_3^{a_1}p_1\inn N\inn\veps_2^{a_2}\,p_3\inn V\inn\veps_3^{a_3}\cI_4\nonumber\\
&&-2p_3^{a_1}p_2\inn V\inn\veps_2^{a_2}\,p_2\inn N\inn\veps_3^{a_3}\cJ_2+2p_3^{a_1}p_2\inn V\inn\veps_2^{a_2}\,p_2\inn V\inn\veps_3^{a_3}\cJ_1-4p_3^{a_1}p_2\inn V\inn\veps_2^{a_2}\,p_3\inn V\inn\veps_3^{a_3}\cJ_3  \nonumber\\
&&+\frac{1}{2}p_3\inn N\inn\veps_2^{a_2}\veps_3^{a_1a_3}\left(\frac{}{}\cJ_{15}p_2\inn N\inn p_3-(2\cJ-\cJ_{16} )p_2\inn V\inn p_3\right) +p_1\inn N\inn\veps_2^{a_2}\veps_3^{a_1a_3}p_3\inn V\inn p_3\cI_4\nonumber\\
&&+  \frac{1}{4}p_3\inn V\inn\veps_2^{a_2}\veps_3^{a_1a_3}\bigg(2p_2\inn V\inn p_2\cJ_1+2p_3\inn V\inn p_3\cJ_4-4p_2\inn N\inn p_3\cJ\bigg)\bigg] +(2\leftrightarrow 3)\labell{A'0A} 
\eeqa 
where $\cJ_{15}=\cJ_{13}-\cJ_{14}$ and $\cJ_{16}=\cJ_{13}+\cJ_{14}$. In above amplitude  $\cJ,\cJ_1,\cJ_2,\cJ_3,\cJ_4,\cJ_5,\cJ_{12},\cJ_{13}$ and $\cJ_{14}$ are some  integrals whose  explicit forms are given in \cite{Garousi:2011ut}. The symmetries of the integrals under the interchange of $(2\leftrightarrow 3)$ are such that $\cJ,\cJ_3,\cJ_{13},\cJ_{14}$ are invariant,  $\cJ_{1}\leftrightarrow\cJ_4 $,  $\cJ_{2}\leftrightarrow\cJ_{12}$, and $\cJ_{5}\leftrightarrow-\cJ_5 $. The terms in the first two lines which are proportional to the integral $\cJ$ are the amplitude for the symmetric NSNS polarization tensors. This part satisfies the Ward identity corresponding to the NSNS gauge transformation. In fact the low energy limit of this part at order $\alpha'^2$ produces only contact terms which are the couplings $C^{(p-3)}\wedge R\wedge R$ in the momentum space \cite{Craps:1998fn,Garousi:2011ut}. The other terms are the amplitude for the antisymmetric NSNS polarization tensors. The combination of these terms and the terms considered in the previous paragraph  satisfy the Ward identity corresponding to the NSNS gauge transformation provided the integrals satisfy the relations
\beqa
-2 \cI_2 p_1\inn N\inn p_2+\cJ_{15}  p_2\inn N\inn p_3+2 \cJ_2 p_2\inn V\inn p_2 +(-4 \cJ+\cJ_{16} -2 \cJ_{5}) p_2\inn V\inn p_3&=&0\nonumber\\
2 \cI_{3} p_1\inn N\inn p_2-2 \cJ_{1} p_2\inn V\inn p_2+\cJ_{15}  p_2\inn V\inn p_3 +(\cJ_{16} +2 \cJ_{5}) p_2\inn N\inn p_3&=&0\nonumber\\
-2 \cI_{4} p_1\inn N\inn p_2+\cJ_{12} p_2\inn N\inn p_3+2 \cJ_{3} p_2\inn V\inn p_2-\cJ_{4} p_2\inn V\inn p_3&=&0\labell{iden3}
\eeqa
and similar relations under the interchange of $(2\leftrightarrow 3)$. One may use these relations and the relation in \reef{id2} to eliminate 8 integrals. Then one left with 6 independent integrals. 
We prefer  to work with the 14 integrals and the 8 constraints.

The amplitude \reef{A0}  does not satisfy the T-dual Ward identity. When the $y$-index is carried by the RR polarization tensor the amplitude is invariant under the linear T-duality. However, when the $y$-index is one of the indices of the NSNS polarization tensors, one finds that the amplitude \reef{A0} is not invariant under the T-duality. The invariance requires the amplitude for the RR $(p-1)$-form with one or two transverse indices. In the next subsection we are going to find such amplitudes by constraining the amplitude to satisfy the T-dual Ward identity, as we have done for the case of two closed and one open string amplitudes in the previous section.  

\subsection{RR $(p-1)$-form}
 
The amplitude for the RR $(p-1)$-form is non-zero when the RR potential has three, two,  one, and zero transverse indices. When the RR potential has one transverse index, the amplitude can be found by applying  the T-dual Ward identity on the amplitude \reef{A0} in which the RR potential has zero transverse index. The T-dual completion of the amplitude \reef{A0} can be written as  
\beqa
{\bf A}_1 &=&{A}_1(f_1)+{\cal A}_1(f_1)\labell{A1}
\eeqa
 where ${\cal A}_1(f_1)$ is the T-dual completion of the amplitude ${\cal A}_0(f_1)$ in \reef{A''0A}, and ${A}_1(f_1)$ is the T-dual completion of the amplitude $A_0(F_1)$ in \reef{A'0A}. The subscribe 1   refers to the number of the transverse index of the RR potential. The T-dual Ward identity on the amplitude \reef{A''0A} requires the following amplitude for ${\cal A}_1(f_1)$:
  \beqa
{\cal A}_1 &\sim& \frac{1}{4}  (f_1 )_{ij}{}^{a_3 }\bigg[\bigg((p_2\inn V\inn\veps_3^S)^{i}(\veps_2^A)^{a_1a_2}-2(p_3\inn V\inn\veps_2^A)^{a_2}(\veps_3^S)^{a_1i}\bigg)(p_2^jp_2^{a_0}\cI_3+p_3^jp_3^{a_0}\cI_2)\nonumber\\
&&\qquad\qquad+\bigg((p_2\inn N\inn\veps_3^S)^{i}(\veps_2^A)^{a_1a_2}-2(p_3\inn N\inn\veps_2^A)^{a_2}(\veps_3^S)^{a_1i}\bigg)(p_3^jp_2^{a_0}\cI_3+p_2^jp_3^{a_0}\cI_2)\nonumber\\
&&\qquad\qquad-4\bigg(p_2^jp_2^{a_0}(p_3\inn V\inn\veps_3^S)^{i}(\veps_2^A)^{a_1a_2}\cI_4-2p_3^j p_3^{a_0}(p_2\inn V\inn\veps_2^A)^{a_2}(\veps_3^S)^{a_1i}\cI_7\bigg)\labell{M1} \\
&&\qquad\qquad+2\bigg(p_2^jp_2^{a_0}(p_1\inn N\inn\veps_3^S)^{i}(\veps_2^A)^{a_1a_2}-2p_3^jp_3^{a_0}(p_1\inn N\inn\veps_2^A)^{a_2}(\veps_3^S)^{a_1i}\bigg)\cI_1+(2\leftrightarrow 3)  \nonumber\\
&& -  \bigg((\veps_3^S)^{a_0i}(\veps_2^A)^{a_1a_2}+(2\leftrightarrow 3)\bigg) \bigg(2p_2^jp_3\inn V\inn p_3\cI_4+p_3^j(p_2\inn V\inn p_3\cI_2-p_2\inn N\inn p_3\cI_3)\bigg)\bigg] \nonumber
 \eeqa
 where  the second $(2\leftrightarrow 3)$ is only for the term in the parenthesis. The RR factor $(f_1)_{ij}{}^{a_3}={p_1}_i {\veps_1}_j{}^{a_3}-{p_1}_j {\veps_1}_i{}^{a_3}$.  The above amplitude  does not satisfy the Ward identity corresponding to the NSNS  and the RR  gauge transformations.

The asymmetry  under the NSNS transformation indicates that the T-dual Ward identity could not capture   terms of the scattering amplitude of the RR $(p-1)$-form which have $(f_1 )^{ija_3}p_{2i}$ or $(f_1)^{ija_3}p_{3i}$. Since all terms in \reef{A''0A} have either $p_{2i}$ or $p_{3i}$, one finds that the missing terms   should have the factor $(f_1 )^{ija_3}p_{2i}p_{3j}$.   Considering all  such terms which have  one   momentum and the two NSNS polarization tensors, with unknown coefficients and imposing the condition that when they combine with the amplitude \reef{M1} they should satisfy the NSNS Ward identity, one finds the following result:
\beqa
{\cal A'}_1  &\sim&\frac{1}{4}p_2^ip_3^j (f_1 )_{ij}{}^{a_3 } \bigg[2p_3^{a_0}\cI_4(\veps_2^A)^{a_1a_2}\Tr[\veps_3^S\inn V]+ \cI_2(p_2\inn N\inn\veps_3^S)^{a_2}(\veps_2^A)^{a_0a_1}\nonumber\\
&&\qquad\qquad\qquad-\cI_3(p_2\inn V\inn\veps_3^S)^{a_2}(\veps_2^A)^{a_0a_1} -(2\leftrightarrow 3)-2 \cI_1(p_1\inn N\inn\veps_3^S)^{a_2}(\veps_2^A)^{a_0a_1}\nonumber\\
&&\qquad\qquad\qquad+2p_3^{a_0}\cG(\veps_2^A\inn V\inn\veps_3^S)^{a_1a_2}\cI_3+2p_3^{a_0}\cG(\veps_3^A\inn V\inn\veps_2^S)^{a_1a_2}\cI_2\bigg]\labell{M1add} 
\eeqa
where the operator $\cG$ which appears in the above amplitude and in the subsequent amplitudes, is defined as
\beqa
{\cal G} (\veps_n^A\inn V\inn\veps_m^S)^{\mu \nu}&\rightarrow &(\veps_n^A\inn V\inn\veps_m^S)^{\mu \nu}-(\veps_n^S\inn N\inn\veps_m^A)^{\mu \nu}\labell{G}\\
{\cal G} (\veps_n^S\inn V\inn\veps_m^A)^{\mu \nu}&\rightarrow &(\veps_n^S\inn V\inn\veps_m^A)^{\mu \nu}-(\veps_n^A\inn N\inn\veps_m^S)^{\mu \nu}\nonumber\\
{\cal G} (\veps_n^A\inn N\inn\veps_m^S)^{\mu \nu}&\rightarrow &(\veps_n^A\inn N\inn\veps_m^S)^{\mu \nu}-(\veps_n^S\inn V\inn\veps_m^A)^{\mu \nu}\nonumber\\
{\cal G} (\veps_n^S\inn N\inn\veps_m^A)^{\mu \nu}&\rightarrow &(\veps_n^S\inn N\inn\veps_m^A)^{\mu \nu}-(\veps_n^A\inn V\inn\veps_m^S)^{\mu \nu}\nonumber
\eeqa
where $n,m$ are the particle labels of the polarization tensors.  The right hand side expressions are invariant  under the linear T-duality when $\mu,\nu\neq y$. One may multiply each term with a momentum, \eg
\beqa
{\cal G} (p_3\inn \veps_2^A\inn V \inn \veps_3^S)^{\mu}=(p_3\inn \veps_2^A\inn V \inn \veps_3^S)^{\mu}-(p_3\inn \veps_2^S\inn N \inn \veps_3^A)^{\mu} 
\eeqa
The result is still invariant under the linear T-duality. 

The combination of the amplitudes \reef{M1} and \reef{M1add} satisfies the NSNS Ward identity, however, they do not satisfy the RR Ward identity. If one includes the amplitude of the RR $(p-1)$-form with two transverse indices which has been found in \cite{Velni:2012sv}, then the RR factor in the above amplitude is extended to the RR field strength $(F_1 )_{ij}{}^{a_3 }={p_1}_i {\veps_1}_j{}^{a_3}-{p_1}_j {\veps_1}_i{}^{a_3}+p_1^{a_3}\veps_{1ij}$, \ie the amplitudes ${\cal A}_1(C_i)+{\cal A'}_1(C_i)$ is extended to ${\cal A}_2(F_{ij})+{\cal A'}_2(F_{ij})$. The amplitude of the RR $(p-1)$-form with two transverse indices has also some terms which become RR gauge invariant after including the amplitude of the RR $(p-1)$-form with three transverse indices \cite{Velni:2012sv}.

The amplitude ${A}_1(f_1)$ in \reef{A1} can be found from imposing the invariance of the amplitude \reef{A'0A} under the linear T-duality when the Killing index $y$ is carried by the NSNS   polarization tensors in \reef{A'0A}. The result is the following: 
\beqa
{A}_1 &\sim& -\frac{1}{2} (f_1 )_i{}^{a_0a_3 }\bigg[p_3^{a_1}\bigg(-p_2\inn N\inn p_3\cG(\veps_2^S\inn V\inn\veps_3^A)^{ia_2} -p_2\inn V\inn p_3\cG(\veps_2^A\inn V\inn\veps_3^S)^{ia_2} \nonumber\\
&&+(p_3\inn N\inn\veps_2^A)^{i}\,(p_2\inn N\inn\veps_3^S)^{a_2}-(p_3\inn V\inn\veps_2^A)^{i}\,(p_2\inn V\inn\veps_3^S)^{a_2}\bigg)\cJ\nonumber\\
&&-(\veps_3^S)^{a_2i}\bigg(p_3^{a_1}(p_3\inn V\inn\veps_2^A\inn V\inn p_2\cJ_1-\frac{1}{2}p_3\inn V\inn\veps_2^A\inn N\inn p_1\cI_3\nonumber\\
&&+p_2\inn V\inn\veps_2^A\inn N\inn p_3\cJ_2 -p_3\inn V\inn\veps_2^A\inn N\inn p_3(\cJ+\cJ_5)+\frac{1}{2}p_3\inn N\inn\veps_2^A\inn N\inn p_1\cI_2)\nonumber\\
&&+\frac{1}{2}(p_3\inn V\inn\veps_2^A)^{a_1}\left(p_2\inn V\inn p_2\cJ_1+p_3\inn V\inn p_3\cJ_4-2p_2\inn N\inn p_3\cJ\right)-2(p_2\inn V\inn\veps_2^A)^{a_1}p_3\inn V\inn p_3\cJ_3\nonumber\\
&&+\frac{1}{2}(p_3\inn N\inn\veps_2^A)^{a_1}\left(p_2\inn N\inn p_3\cJ_{15}+p_2\inn V\inn p_3(\cJ_{16} -2\cJ)\right)+(p_1\inn N\inn\veps_2^A)^{a_1}p_3\inn V\inn p_3\cI_4 \bigg)\nonumber\\
&&+(p_1\inn N\inn\veps_3^S)^{i}\bigg(p_3^{a_1}((p_1\inn N\inn\veps_2^A)^{a_2}\cI_1+\frac{1}{2}(p_3\inn V\inn\veps_2^A)^{a_2}\cI_2\nonumber\\
&&-\frac{1}{2}(p_3\inn N\inn\veps_2^A)^{a_2}\cI_3 -2 (p_2\inn V\inn\veps_2^A)^{a_2}\cI_7)-\frac{1}{2}(\veps_2^A)^{a_1a_2}p_2\inn V\inn p_2\cI_7\bigg)+\nonumber\\
&&+(p_2\inn N\inn\veps_3^S)^{i}\bigg(p_3^{a_1}(-\frac{1}{2}(p_1\inn N\inn\veps_2^A)^{a_2}\cI_2 -(p_3\inn V\inn\veps_2^A)^{a_2}\cJ_5 +(p_2\inn V\inn\veps_2^A)^{a_2}\cJ_2)\nonumber\\
&&-\frac{1}{4}(\veps_2^A)^{a_1a_2}(p_2\inn N\inn p_3\cJ_{15}+p_2\inn V\inn p_3(\cJ_{16} -2\cJ))\bigg)\nonumber\\
&&+(p_2\inn V\inn\veps_3^S)^{i}\bigg(p_3^{a_1}(\frac{1}{2}(p_1\inn N\inn\veps_2^A)^{a_2}\cI_3 -(p_2\inn V\inn\veps_2^A)^{a_2}\cJ_1 +(p_3\inn N\inn\veps_2^A)^{a_2}\cJ_5)\nonumber\\
&&-\frac{1}{4}(\veps_2^A)^{a_1a_2}(p_2\inn V\inn p_2\cJ_1+p_3\inn V\inn p_3\cJ_4-2p_2\inn N\inn p_3\cJ)\bigg)\nonumber\\
&&+(p_3\inn V\inn\veps_3^S)^{i}\bigg(p_3^{a_1}(-2(p_1\inn N\inn\veps_2^A)^{a_2}\cI_4 +4 (p_2\inn V\inn\veps_2^A)^{a_2}\cJ_3 +(p_3\inn N\inn\veps_2^A)^{a_2}\cJ_{12}\nonumber\\
&&-(p_3\inn V\inn\veps_2^A)^{a_2}\cJ_4)+(\veps_2^A)^{a_1a_2}p_2\inn V\inn p_2\cJ_3\bigg)\bigg]+(2\leftrightarrow 3)\labell{A'1} 
\eeqa
 where  the RR factor is $(f_1)_i{}^{a_0a_3}={p_1}^{a_3} {\veps_1}_i{}^{a_0}-{p_1}^{a_0} {\veps_1}_i{}^{a_3}$ . The above amplitude is invariant under the linear T-duality when the world volume $y$-index is carried by the RR potential. Note that the two NSNS polarization tensors in the first line contract with each other  in such a way that they are invariant under the T-duality when $a_2\neq y$. The above amplitude does not satisfy the Ward identity corresponding to the NSNS and the RR gauge transformations.

The asymmetry  under the NSNS gauge transformation indicates that the T-dual Ward identity could not capture   terms of the scattering amplitude of the RR $(p-1)$-form which have $(f_1 )^{ia_0a_3}p_{2i}$ or $(f_1)^{ia_0a_3}p_{3i}$. Since no term in \reef{A'0A} has  $p_{2i}$ or $p_{3i}$, one finds that there are two types of  missing terms. One type is the terms which have    $(f_1 )^{ia_0a_3}p_{2i}$, and the other type is the terms  which  have  $(f_1)^{ia_0a_3}p_{3i}$. Therefore, the missing terms can be separated as $A'_1=A'_{12}+A'_{13}$.  Considering all  such terms which have  two   momentum and the two NSNS polarization tensors, with unknown coefficients and imposing the condition that when they combine with the amplitude \reef{A'1} they should satisfy the NSNS Ward identity, one finds the following result for the terms of the first type:
\beqa
{A'}_{12} &\!\!\!\!\!\!\!\sim\!\!\!\!\!\!\!& -\frac{1}{4}(f_1 )^{ia_0a_3}p_{2i}\bigg[p_3^{a_1}\bigg(\cG(p_2\inn N\inn\veps_3^S\inn N\inn\veps_2^A)^{a_2}(\cJ_{16} -2\cJ_5)+2\cG(p_1\inn N\inn\veps_3^S\inn N\inn\veps_2^A)^{a_2}\cI_{2}\nonumber\\
&&+\cG(p_2\inn V\inn\veps_3^S\inn V\inn\veps_2^A)^{a_2}(\cJ_{16} -4\cJ+2\cJ_5)+4 \cG(p_3\inn V\inn\veps_3^S\inn V\inn\veps_2^A)^{a_2}\cJ_{12}\nonumber\\
&&+(\cG(p_2\inn V\inn\veps_3^S\inn N\inn\veps_2^A)^{a_2}+\cG(p_2\inn N\inn\veps_3^S\inn V\inn\veps_2^A)^{a_2})\cJ_{15}-2\cG(p_1\inn N\inn\veps_3^S\inn V\inn\veps_2^A)^{a_2}\cI_{3}\nonumber\\
&&-4\cG(p_3\inn V\inn\veps_3^S\inn N\inn\veps_2^A)^{a_2}\cJ_{4}-2p_3\inn V\inn p_3\cG(\veps_2^A\inn V\inn\veps_3^S)^{a_1a_2}\cJ_{12}\bigg)\nonumber\\
&&+\frac{1}{2}(\veps_2^A)^{a_1a_2}\bigg(p_2\inn N\inn\veps_3^S\inn N\inn p_2(\cJ_{16} -2\cJ_5)-4\cJ_{4} p_3\inn V\inn\veps_3^S\inn N\inn p_2+4\cI_2p_2\inn N\inn\veps_3^S\inn N\inn p_1\nonumber\\
&&+p_2\inn V\inn\veps_3^S\inn V\inn p_2(\cJ_{16}-4\cJ +2\cJ_5)+4p_3\inn V\inn\veps_3^S\inn V\inn p_2\cJ_{12}-2 p_2\inn V\inn\veps_3^S\inn N\inn p_1\cI_3\nonumber\\
&&+2p_2\inn V\inn\veps_3^S\inn N\inn p_2\cJ_{15}+2(-2\cI_{4}p_1\inn N\inn p_3+\cJ_4p_2\inn N\inn p_3-\cJ_{12}p_2\inn V\inn p_3)\Tr[\veps_3^S\inn V]\bigg)\nonumber\\
&&+ (p_3\inn N\inn\veps_2^A)^{a_2}\bigg(2p_3^{a_1}\Tr[\veps_3^S\inn V]\cJ_{4}-2 \cI_{2}(p_1\inn N\inn\veps_3^S)^{a_1}-(p_2\inn N\inn\veps_3^S)^{a_1}\,(\cJ_{16} -2\cJ_5)\nonumber\\
&&-(p_2\inn V\inn\veps_3^S)^{a_1}\cJ_{15}\bigg)-(p_3\inn V\inn\veps_2^A)^{a_2}\bigg(2p_3^{a_1} \Tr[\veps_3^S\inn V]\cJ_{12}+(p_2\inn N\inn\veps_3^S)^{a_1}\cJ_{15}\nonumber\\
&&+(p_2\inn V\inn\veps_3^S)^{a_1}\,(\cJ_{16}-4\cJ +2\cJ_5)\bigg)\bigg] +(2\leftrightarrow 3)\labell{M'1add} 
\eeqa
The terms of the second type which have the RR factor  $(f_1)^{ia_0a_3}p_{3i}$ are the following:
\beqa
{A'}_{13} &\!\!\!\!\!\!\!\sim\!\!\!\!\!\!\!&- \frac{1}{4} (f_1)^{ia_0a_3}p_{3i} \bigg[p_3^{a_1}\bigg(\cG(p_2\inn V\inn\veps_3^S\inn N\inn\veps_2^A)^{a_2}(\cJ_{16} -2\cJ_5)+\cG(p_2\inn N\inn\veps_3^S\inn V\inn\veps_2^A)^{a_2}\nonumber\\
&&\times(\cJ_{16}-4\cJ +2\cJ_5)+(\cG(p_2\inn V\inn\veps_3^S\inn V\inn\veps_2^A)^{a_2} +\cG (p_2\inn N\inn\veps_3^S\inn N\inn\veps_2^A)^{a_2})\cJ_{15}\bigg)\nonumber\\
&&+\frac{1}{2}(\veps_2^A)^{a_1a_2}\bigg(4(p_1\inn N\inn p_2\cI_{4}-p_2\inn V\inn p_2\cJ_{3})\Tr[\veps_3^S\inn V]+2p_2\inn V\inn\veps_3^S\inn N\inn p_2\,(\cJ_{16}-2\cJ )\nonumber\\
&&+(p_2\inn N\inn\veps_3^S\inn N\inn p_2\,+p_2\inn V\inn\veps_3^S\inn V\inn p_2)\cJ_{15}+2p_2\inn V\inn\veps_3^S\inn N\inn p_1\, \cI_2\bigg)\labell{M''1add} \\
&&+(p_3\inn V\inn\veps_2^A)^{a_2}\bigg(2p_3^{a_1}\Tr[\veps_3^S\inn V] \cJ_{4}-2(p_1\inn N\inn\veps_3^S)^{a_1}\cI_2-(p_2\inn V\inn \veps_3^S)^{a_1}\cJ_{15}\nonumber\\
&& -(p_2\inn N\inn\veps_3^S)^{a_1}(\cJ_{16} -4\cJ-2\cJ_5)\bigg)+2(p_1\inn N\inn\veps_2^A)^{a_2}\bigg(2p_3^{a_1}\Tr[\veps_3^S\inn V] \cI_{4}\nonumber\\
&&-2 (p_1\inn N\inn\veps_3^S)^{a_1}\cI_{1}+(p_2\inn N\inn\veps_3^S)^{a_1}\cI_{2}-(p_2\inn V\inn\veps_3^S)^{a_1}\cI_{3}\bigg)-4(p_2\inn V\inn\veps_2^A)^{a_2}\nonumber\\
&&\times\bigg((p_2\inn N\inn\veps_3^S)^{a_1}\cJ_{2}-2 (p_1\inn N\inn\veps_3^S)^{a_1}\cI_{7}-(p_2\inn V\inn\veps_3^S)^{a_1}\cJ_{1}+2p_3^{a_1}\Tr[\veps_3^S\inn V] \cJ_{3}\bigg)\nonumber\\
&&-(p_3\inn N\inn\veps_2^A)^{a_2}\bigg( (p_2\inn N\inn\veps_3^S)^{a_1}\cJ_{15}+(p_2\inn V\inn\veps_3^S)^{a_1}(\cJ_{16} +2\cJ_5)+2p_3^{a_1}\Tr[\veps_3^S\inn V] \cJ_{12}\bigg)\nonumber\\
&&-2p_2\inn V\inn p_2  \cG(\veps_2^A\inn V\inn\veps_3^S)^{a_1a_2}\cJ_1 +(2\leftrightarrow 3)+2p_1\inn N\inn p_2 \bigg(\cG(\veps_2^A\inn V\inn\veps_3^S)^{a_1a_2}\cI_3+(2\leftrightarrow 3)\bigg) \bigg]\nonumber
\eeqa
The coefficient of the first term in the second line of \reef{M'1add} and  the coefficient of the last term in the first line of \reef{M''1add} are not fixed by the condition that the sum of the above amplitudes and the amplitude \reef{A'1} to be invariant under the NSNS gauge transformation. That coefficients will be fixed in the next section by using the appropriate Ward identity.  In above equations we have written the final result.

The sum of  the amplitudes \reef{A'1}, \reef{M'1add} and \reef{M''1add} satisfies the NSNS Ward identities.  However, it does not satisfy the RR Ward identity because the RR factor $(f_1)^{ia_0a_3}$ is not the RR field strength. It can easily be extended to the RR invariant amplitude by extending the RR factor to the RR field strength $(F_1)^{ia_0a_3}=p_1^i\veps_1^{a_0a_3}+p_1^{a_3}\veps_1^{ia_0}-p_1^{a_0}\veps_1^{ia_3 }$. The amplitude corresponding to the first term does not satisfy the NSNS Ward identity. So one has to add the  amplitude of the  RR $(p-1)$-form with no transverse index. The RR invariance requires this  amplitude to be in terms of the RR field strength $(F_1)^{a_0a_1a_3}$.  One may   consider  all independent terms containing  $(F_1)^{a_0a_1a_3}$ which are   196 terms, and then impose the condition that when they combine with the above non-gauge invariant amplitude, the combination satisfies the NSNS Ward identity. We have done this calculation and found that the NSNS Ward identity  fixes  171 unknown integrals and the 25 remaining integrals  appear in some constraint equations, like the constraint in \reef{2cons}. However, to analyze the amplitude at low energy one needs the explicit form of the integrals which   may be found by performing the    string theory calculations to find the amplitude of the  RR $(p-1)$-form   which have structure $(F_1)^{a_0a_1a_3}(\cdots)^{a_2}$. We leave the details of theses calculation for the future work.


\subsection{RR $(p+1)$-form}

The amplitude for the RR $(p+1)$-form is non-zero when the RR potential has four, three, two,  and one transverse indices.  When the RR potential has two transverse indices, the amplitude can be found by applying  the T-dual Ward identity on the amplitude \reef{A1}, \ie
\beqa
{\bf A}_2 &=&{A}_2(f_1)+{\cal A}_2(f_1)\labell{A2}
\eeqa
 The amplitude  ${\cal A}_2(f_1)$ is the T-dual completion of the amplitude ${\cal A}_1(f_1)$ in \reef{M1}, and ${A}_2(f_1)$ is the T-dual completion of the amplitude $A_1(f_1)$ in \reef{A'1}.  The   amplitude  ${\cal A}_2(f_1)$ is 
  \beqa
{\cal A}_2&\sim&-\frac{1}{4}(f _1)_{ijk}{}^{a_2 }\bigg[\bigg(2(p_3\inn V\inn\veps_2^S)^{i}(\veps_3^S)^{a_1j}+(p_3\inn V\inn\veps_2^A)^{a_1}(\veps_3^A)^{ij}\bigg)(p_2^kp_2^{a_0}\cI_3+p_3^kp_3^{a_0}\cI_2)\nonumber\\
&&+\bigg(2(p_3\inn N\inn\veps_2^S)^{i}(\veps_3^S)^{a_1j}+(p_3\inn N\inn\veps_2^A)^{a_1}(\veps_3^A)^{ij}\bigg)(p_3^kp_2^{a_0}\cI_3+p_2^kp_3^{a_0}\cI_2)\nonumber\\
&&-4p_3^kp_3^{a_0}\bigg(2(p_2\inn V\inn\veps_2^S)^{i}(\veps_3^S)^{a_1j}+(p_2\inn V\inn\veps_2^A)^{a_1}(\veps_3^A)^{ij}\bigg)\cI_7\nonumber\\
&&+2p_3^kp_3^{a_0}\bigg(2(p_1\inn N\inn\veps_2^S)^{i}(\veps_3^S)^{a_1j}+(p_1\inn N\inn\veps_2^A)^{a_1}(\veps_3^A)^{ij}\bigg)\cI_1 +(2\leftrightarrow 3)\nonumber\\
&&+\frac{1}{2}\bigg((\veps_2^A)^{ij}(\veps_3^A)^{a_0a_1}+2(\veps_2^S)^{a_0i}(\veps_3^S)^{a_1j}+(2\leftrightarrow 3)\bigg)\nonumber\\
&&\times\bigg(2p_2^kp_3\inn V\inn p_3\cI_4+p_3^kp_2\inn V\inn p_3\cI_2-p_3^kp_2\inn N\inn p_3\cI_3\bigg)\bigg]\labell{M2}
\eeqa
where the RR factor is $(f_1)^{ijka_2}=p_1^i\veps_1^{jka_2}+p_1^k\veps_1^{ija_2}+p_1^j\veps_1^{kia_2}$. For simplicity we have considered the amplitude for $p=2$. The amplitudes in the previous section is nonzero when one of the NSNS polarization tensors is symmetric and the other one is antisymmetric. The amplitudes in this section which are the T-dual completion of the amplitude in the previous section  are then non-zero when both tensors are symmetric or both are antisymmetric. The above amplitude does not satisfy the Ward identity corresponding to the NSNS  and the RR  gauge transformations.

The asymmetry  under the NSNS transformation indicates that the T-dual Ward identity could not capture   terms of the scattering amplitude of the RR $(p+1)$ which have $(f_1 )^{ijka_2}p_{2i}$ or $(f_1)^{ijka_2}p_{3i}$. Since all terms in \reef{M1} have either $p_{2i}$ or $p_{3i}$, one finds that the missing terms  corresponding to the above amplitude  should have the factor $(f_1 )^{ijka_2}p_{2i}p_{3j}$.   Considering all  such terms which have  one   momentum and the two NSNS polarization tensors, with unknown coefficients and imposing the condition that when they combine with the amplitude \reef{M2} they should satisfy the NSNS Ward identity, one finds the following result:
\beqa
{\cal A'}_2  &\!\!\!\!\!\!\!\!\!\!\sim\!\!\!\!\!\!\!\!\!\!&\frac{1}{4} p_2^ip_3^j (f_1)_{ijk}{}^{a_2}\bigg[4p_3^{a_0}\cI_4(\veps_2^S)^{a_1k}\Tr[\veps_3^S\inn V]+\cI_2\bigg((p_2\inn N\inn\veps_3^A)^{k}(\veps_2^A)^{a_0a_1}-2 (p_2\inn N\inn\veps_3^S)^{a_1}(\veps_2^S)^{a_0k}\bigg)\nonumber\\
&&+\cI_3\bigg(2(p_2\inn V\inn\veps_3^S)^{a_1}(\veps_2^S)^{a_0k}-(p_2\inn V\inn\veps_3^A)^{k}(\veps_2^A)^{a_0a_1}\bigg)-(2\leftrightarrow 3)+2p_3^{a_0}\cG\bigg((\veps_2^S\inn V\inn\veps_3^S)^{ka_1}\cI_3\nonumber\\
&&+(\veps_2^A\inn V\inn\veps_3^A)^{a_1k}\cI_3+(2\leftrightarrow 3)\bigg)-2 \cI_1\bigg((p_1\inn N\inn\veps_3^A)^{k}(\veps_2^A)^{a_0a_1}-2(p_1\inn N\inn\veps_3^S)^{a_1}(\veps_2^S)^{a_0k}\bigg)\bigg]\labell{M2add}
\eeqa
The above amplitude is also the T-dual completion of the amplitude \reef{M1add}. In fact the amplitude \reef{M1add} is invariant/covariant under the linear T-duality when the world volume Killing $y$-index is carried by the RR potential. However, when the world volume $y$-index is carried by the NSNS polarization tensors in \reef{M1add}, the amplitude does not transform to itself under the linear T-duality. It produces the above amplitude for $k=y$ under T-duality. Completing the transverse $y$-index, one finds the above amplitude. 

The combination of the amplitudes \reef{M2} and \reef{M2add} satisfies the NSNS Ward identity, however, they do not satisfy the RR Ward identity. If one includes the amplitude of the RR $(p+1)$-form with three transverse indices which has been found in \cite{Velni:2012sv}, then the RR factor in the above amplitude is extended to the RR field strength $(F_1)^{ijka_2}=p_1^i\veps_1^{jka_2}+p_1^k\veps_1^{ija_2}+p_1^j\veps_1^{kia_2}-p_1^{a_2}\veps_1^{ijk}$,  \ie the amplitudes ${\cal A}_2(C_{ij})+{\cal A'}_2(C_{ij})$ is extended to ${\cal A}_3(F_{ijk})+{\cal A'}_3(F_{ijk})$. The amplitude of the RR $(p+1)$-form with three transverse indices has also some terms which become RR gauge invariant after combining them with  the amplitude of the RR $(p+1)$-form with four transverse indices \cite{Velni:2012sv}. However, the RR invariant amplitude does not satisfy the NSNS Ward identity anymore . Here also,  unlike the amplitudes  ${\cal A}_2(C_{ij})+{\cal A'}_2(C_{ij})$ which satisfy the NSNS Ward identity but do not satisfy the RR Ward identity, the amplitudes ${\cal A}_3(F_{ijk})+{\cal A'}_3(F_{ijk})$ do not satisfy the NSNS Ward identity but satisfy the RR Ward identity. To make the amplitudes ${\cal A}_3(F_{ijk})+{\cal A'}_3(F_{ijk})$ to satisfy the NSNS Ward identity, one requires to take into account  the other amplitudes for the RR $(p+1)$-form which we are going to consider now. 

The amplitude ${A}_2(f_1)$ in \reef{A2} can be found by imposing the invariance of the amplitude \reef{A'1} under the linear T-duality. The amplitude \reef{A'1} is invariant under the linear T-duality when the world volume index $y$ is contracted with the RR potential, however,  when the Killing index $y$ is contracted with the NSNS   polarization tensors in \reef{A'1}, \ie $a_1=y$ or $a_2=y$, then the amplitude produces new terms under the linear T-duality. Completing the transverse $y$-index in the new terms, one finds  the following result:
\beqa
{A}_2 &\sim&  \frac{1}{2}(f_1)_{ij}{}^{a_0a_2}\bigg[p_3^{a_1}\bigg(-p_2\inn N\inn p_3\cG (\veps_2^S\inn V\inn\veps_3^S)^{ij}-p_2\inn V\inn p_3\cG(\veps_2^A\inn V\inn\veps_3^A)^{ij}\nonumber\\
&&+(p_3\inn N\inn\veps_2^A)^{i}\,(p_2\inn N\inn\veps_3^A)^{j}-(p_3\inn V\inn\veps_2^A)^{i}(p_2\inn V\inn\veps_3^A)^{j} \bigg)\cJ\nonumber\\
&&-(\veps_3^A)^{ij}\bigg(p_3^{a_1}(p_3\inn V\inn\veps_2^A\inn V\inn p_2\cJ_1-\frac{1}{2}p_3\inn V\inn\veps_2^A\inn N\inn p_1\cI_3+p_2\inn V\inn\veps_2^A\inn N\inn p_3\cJ_2\nonumber\\
&&-p_3\inn V\inn\veps_2^A\inn N\inn p_3(\cJ+\cJ_5)+\frac{1}{2}p_3\inn N\inn\veps_2^A\inn N\inn p_1\cI_2)+\frac{1}{2}(p_3\inn V\inn\veps_2^A)^{a_1}(p_2\inn V\inn p_2\cJ_1\nonumber\\
&&+p_3\inn V\inn p_3\cJ_4-2p_2\inn N\inn p_3\cJ)+(p_1\inn N\inn\veps_2^A)^{a_1}p_3\inn V\inn p_3\cI_4 -(p_2\inn V\inn\veps_2^A)^{a_1} p_3\inn V\inn p_3\cJ_3\nonumber\\
&&+\frac{1}{2}(p_3\inn N\inn\veps_2^A)^{a_1}(\cJ_{15}p_2\inn N\inn p_3+(\cJ_{16}-2\cJ)p_2\inn V\inn p_3)\bigg)\nonumber\\
&&-(p_1\inn N\inn\veps_2^S)^{i}\bigg(p_3^{a_1}((p_2\inn V\inn\veps_3^S)^{j}\cI_3 +(p_1\inn N\inn\veps_3^S)^{j}\cI_1-(p_2\inn N\inn\veps_3^S)^{j}\cI_2 - 4(p_3\inn V\inn\veps_3^S)^{j}\cI_4)\nonumber\\
&&+2(\veps_3^S)^{a_1j}p_3\inn V\inn p_3\cI_4\bigg)-2(p_2\inn V\inn\veps_2^S)^{i}\bigg(p_3^{a_1}((p_2\inn N\inn\veps_3^S)^{j}\cJ_2\labell{M'2}\\
&&-(p_2\inn V\inn\veps_3^S)^{j}\cJ_1+2 (p_3\inn V\inn\veps_3^S)^{j}\cJ_3)-2(\veps_3^S)^{a_1j}p_3\inn V\inn p_3\cJ_3\bigg)\nonumber\\
&&+(p_3\inn V\inn\veps_2^S)^{i}\bigg(2p_3^{a_1}(p_2\inn N\inn\veps_3^S)^{j}\cJ_5-(\veps_3^S)^{a_1j}(p_2\inn V\inn p_2\cJ_1+p_3\inn V\inn p_3\cJ_4-2p_2\inn N\inn p_3\cJ)\bigg)\nonumber\\
&&-(p_3\inn N\inn\veps_2^S)^{i}(\veps_3^S)^{a_1j}(\cJ_{15}p_2\inn N\inn p_3+(\cJ_{16}-2\cJ)p_2\inn V\inn p_3)\bigg]+(2\leftrightarrow 3)\nonumber
\eeqa 
where the RR factor is $(f_1)^{ij a_0a_2}=p_1^{a_0}\veps_1^{ija_2}-p_1^{a_2}\veps_1^{ija_0}$. The above amplitude does not satisfy the Ward identity corresponding to the NSNS and the RR gauge transformations.

The asymmetry  under the NSNS gauge transformation indicates that the T-dual Ward identity could not capture   terms of the scattering amplitude of the RR $(p+1)$ which have $(f_1 )^{ija_0a_2}p_{2i}$ or $(f_1)^{ija_0a_2}p_{3i}$. Since the RR factor carries two transverse indices, one finds that there are three types of  missing terms in this case. The first type is the terms which have    $(f_1 )^{ija_0a_2}p_{2i}$,  the second type is the terms  which  have  $(f_1)^{ija_0a_2}p_{3i}$  and the third type  is the terms  which  have  $(f_1)^{ija_0a_2}p_{2i}p_{3j}$. None of these terms can be captured by the T-dual Ward identity because the momentum along the $y$-direction is zero in the T-duality transformation. In fact under the T-duality rules one has $(f_1)^{iya_0a_2}p_{2i}p_{3y}=0$ or  $(f_1)^{yja_0a_2}p_{2y}p_{3j}=0$. Therefore, the missing terms can be separated as $A'_2=A'_{22}+A'_{23}+A_2''$.  One may consider all  such terms   with unknown coefficients and impose the NSNS Ward identity to find the coefficients.  Alternatively, one may find the amplitudes $A'_{22}+A'_{23}$ by imposing the T-dual Ward identity on the amplitudes \reef{M'1add} and \reef{M''1add}, and then find the amplitude $A''_2$ from the NSNS Ward identity. Note that the RR factor of $A''_2$ which is $(f_1)^{ija_0a_2}p_{2i}p_{3j}$ does not allow the NSNS polarization tensors to carry the transverse index of the RR factor. As a result, these terms can not be the T-dual completion of the amplitudes in the previous section.  

The   amplitude \reef{M'1add} has the overall RR factor $(f_1 )^{ia_0a_3}p_{2i}$. Therefore, the T-dual completion of this amplitude produces new terms in  the first type. In fact, the amplitude \reef{M'1add} is invariant under the linear T-duality when the world volume index $y$ is contracted with the RR potential, however,  when the Killing index $y$ is contracted with the NSNS   polarization tensors in \reef{M'1add}, \ie $a_1=y$ or $a_2=y$, then the amplitude produces new terms under the linear T-duality. Completing the transverse $y$-index in the new terms, one finds  the following result:
\beqa
{A'}_{22}&\!\!\!\!\!\!\!\!\!\!\sim\!\!\!\!\!\!\!\!\!\!&-\frac{1}{4}(f_1 )_{ij}{}^{a_0a_2}p_{2}^i\bigg[-2p_3\inn V\inn p_3\bigg(\cG(\veps_2^S\inn V\inn\veps_3^S)^{ja_1}+\cG(\veps_2^A\inn V\inn\veps_3^A)^{a_1j}\bigg)\cJ_{12}\labell{M'2add}\\
&&+p_3^{a_1}\bigg(\cG(p_2\inn N\inn\veps_3^S\inn N\inn\veps_2^S)^{j}(\cJ_{16}-2\cJ_5)+\cG(p_2\inn V\inn\veps_3^S\inn V\inn\veps_2^S)^{j}(\cJ_{16}-4\cJ+2\cJ_5)\nonumber\\
&&+4 \cG(p_3\inn V\inn\veps_3^S\inn V\inn\veps_2^S)^{j}\cJ_{12}+(\cG(p_2\inn V\inn\veps_3^S\inn N\inn\veps_2^S)^{j}+\cG(p_2\inn N\inn\veps_3^S\inn V\inn\veps_2^S)^{j})\cJ_{15}\nonumber\\
&&-2\cG(p_1\inn N\inn\veps_3^S\inn V\inn\veps_2^S)^{j}\cI_{3}+2\cG(p_1\inn N\inn\veps_3^S\inn N\inn\veps_2^S)^{j}\cI_{2}-4\cG(p_3\inn V\inn\veps_3^S\inn N\inn\veps_2^S)^{j}\cJ_{4}\bigg)\nonumber\\
&&+ (p_3\inn N\inn\veps_2^A)^{a_1}\bigg(2 \cI_{2}(p_1\inn N\inn\veps_3^A)^{j}+(p_2\inn N\inn\veps_3^A)^{j}(\cJ_{16}-2\cJ_5)+(p_2\inn V\inn\veps_3^A)^{j}\cJ_{15}\bigg)\nonumber\\
&&+ (p_3\inn N\inn\veps_2^S)^{j}\bigg(2p_3^{a_1}\Tr[\veps_3^S\inn V]\cJ_{4}-2 \cI_{2}(p_1\inn N\inn\veps_3^S)^{a_1}-(p_2\inn N\inn\veps_3^S)^{a_1}(\cJ_{16}-2\cJ_5)\nonumber\\
&&-(p_2\inn V\inn\veps_3^S)^{a_1}\cJ_{15}\bigg)+(p_3\inn V\inn\veps_2^A)^{a_1}\bigg((p_2\inn N\inn\veps_3^A)^{j}\cJ_{15}+(p_2\inn V\inn\veps_3^A)^{j}(\cJ_{16}-4\cJ+2\cJ_5)\bigg)\nonumber\\
&&-(p_3\inn V\inn\veps_2^S)^{j}\bigg(2p_3^{a_1} \Tr[\veps_3^S\inn V]\cJ_{12}+(p_2\inn N\inn\veps_3^S)^{a_1}\cJ_{15}+(p_2\inn V\inn\veps_3^S)^{a_1}(\cJ_{16}-4\cJ+2\cJ_5)\bigg)\nonumber\\
&&+(\veps_2^S)^{a_1j}\bigg(2p_2\inn V\inn\veps_3^S\inn N\inn p_2\cJ_{15}-2(2\cI_{4}p_1\inn N\inn p_3-\cJ_4p_2\inn N\inn p_3+\cJ_{12}p_2\inn V\inn p_3)\Tr[\veps_3^S\inn V]\nonumber\\
&&+p_2\inn V\inn\veps_3^S\inn V\inn p_2(\cJ_{16}-4\cJ+2\cJ_5)+4p_3\inn V\inn\veps_3^S\inn V\inn p_2\cJ_{12}-2 p_2\inn V\inn\veps_3^S\inn N\inn p_1\cI_3\nonumber\\
&&+p_2\inn N\inn\veps_3^S\inn N\inn p_2(\cJ_{16}-2\cJ_5)-4\cJ_{4} p_3\inn V\inn\veps_3^S\inn N\inn p_2+4\cI_2p_2\inn N\inn\veps_3^S\inn N\inn p_1\bigg)\bigg]+(2\leftrightarrow 3)\nonumber
\eeqa
This amplitude is invariant under the linear T-duality when the $y$-index is carried by the RR potential, otherwise, it is not invariant. We will consider the T-dual completion of this amplitude in the next section.

The terms of the second type which have the RR factor  $(f_1)^{ija_0a_2}p_{3i}$ can be found from the T-dual completion of the amplitude \reef{M''1add} because this amplitude has the overall RR factor $(f_1 )^{ia_0a_3}p_{3i}$.  Here also one realizes that the amplitude \reef{M''1add} is invariant under the linear T-duality only when the world volume index $y$ is contracted with the RR potential. When  it is contracted with the NSNS   polarization tensors,  the amplitude produces new terms under the linear T-duality which have the transverse $y$-index. Completing this index, one finds  the following result:
\beqa
{A'}_{23}&\!\!\!\!\!\!\!\!\!\!\sim\!\!\!\!\!\!\!\!\!\!& -\frac{1}{4}(f_1)_{ij}{}^{a_0a_2}p_{3}^i\bigg[ -2p_2\inn V\inn p_2 \bigg( \cG(\veps_2^S\inn V\inn\veps_3^S)^{j a_1}+\cG(\veps_2^A\inn V\inn\veps_3^A)^{a_1 j}\bigg)\cJ_1\nonumber\\
&&+p_3^{a_1}\bigg(\cG(p_2\inn V\inn\veps_3^S\inn N\inn\veps_2^S)^{j}(\cJ_{16}-2\cJ_5)+\cG(p_2\inn N\inn\veps_3^S\inn V\inn\veps_2^S)^{j}(\cJ_{16}-4\cJ+2\cJ_5)\nonumber\\
&&+(\cG(p_2\inn V\inn\veps_3^S\inn V\inn\veps_2^S)^{j}+\cG (p_2\inn N\inn\veps_3^S\inn N\inn\veps_2^S)^{j})\cJ_{15}\bigg)+(p_3\inn V\inn\veps_2^A)^{a_1}\bigg(2(p_1\inn N\inn\veps_3^A)^{j}\cI_2\nonumber\\
&&+(p_2\inn N\inn\veps_3^A)^{j}(\cJ_{16}-4\cJ-2\cJ_5)+(p_2\inn V\inn \veps_3^A)^{j}\cJ_{15}\bigg)+(p_3\inn V\inn\veps_2^S)^{j}\bigg(2p_3^{a_1}\Tr[\veps_3^S\inn V] \cJ_{4}\nonumber\\
&&-2(p_1\inn N\inn\veps_3^S)^{a_1}\cI_2 -(p_2\inn N\inn\veps_3^S)^{a_1}(\cJ_{16}-4\cJ-2\cJ_5)-(p_2\inn V\inn \veps_3^S)^{a_1}\cJ_{15}\bigg)\nonumber\\
&&+2(p_1\inn N\inn\veps_2^A)^{a_1}\bigg(2 (p_1\inn N\inn\veps_3^A)^{j}\cI_{1}-(p_2\inn N\inn\veps_3^A)^{j}\cI_{2}+(p_2\inn V\inn\veps_3^A)^{j}\cI_{3}\bigg)\nonumber\\
&&+2(p_1\inn N\inn\veps_2^S)^{j}\bigg(2p_3^{a_1}\Tr[\veps_3^S\inn V] \cI_{4}-2 (p_1\inn N\inn\veps_3^S)^{a_1}\cI_{1}+(p_2\inn N\inn\veps_3^S)^{a_1}\cI_{2}-(p_2\inn V\inn\veps_3^S)^{a_1}\cI_{3}\bigg)\nonumber\\
&&-4(p_2\inn V\inn\veps_2^A)^{a_1}\bigg(2 (p_1\inn N\inn\veps_3^A)^{j}\cI_{7}-(p_2\inn N\inn\veps_3^A)^{j}\cJ_{2}+(p_2\inn V\inn\veps_3^A)^{j}\cJ_{1}\bigg)\nonumber\\
&&-4(p_2\inn V\inn\veps_2^S)^{j}\bigg(2p_3^{a_1}\Tr[\veps_3^S\inn V] \cJ_{3}-2 (p_1\inn N\inn\veps_3^S)^{a_1}\cI_{7}+(p_2\inn N\inn\veps_3^S)^{a_1}\cJ_{2}-(p_2\inn V\inn\veps_3^S)^{a_1}\cJ_{1}\bigg)\nonumber\\
&&+(p_3\inn N\inn\veps_2^A)^{a_1}\bigg((p_2\inn N\inn\veps_3^A)^{j}\cJ_{15}+(p_2\inn V\inn\veps_3^A)^{j}(\cJ_{16}+2\cJ_5)\bigg)\nonumber\\
&&-(p_3\inn N\inn\veps_2^S)^{j}\bigg(2p_3^{a_1}\Tr[\veps_3^S\inn V] \cJ_{12}+(p_2\inn N\inn\veps_3^S)^{a_1}\cJ_{15}+(p_2\inn V\inn\veps_3^S)^{a_1}(\cJ_{16}+2\cJ_5)\bigg)\nonumber\\
&&+(\veps_2^S)^{a_1j}\bigg(4(p_1\inn N\inn p_2\cI_{4}-p_2\inn V\inn p_2\cJ_{3})\Tr[\veps_3^S\inn V]+2p_2\inn V\inn\veps_3^S\inn N\inn p_2\,(\cJ_{16}-2\cJ)\nonumber\\
&&+(p_2\inn N\inn\veps_3^S\inn N\inn p_2\,+p_2\inn V\inn\veps_3^S\inn V\inn p_2)\cJ_{15}+2p_2\inn V\inn\veps_3^S\inn N\inn p_1\, \cI_2\bigg) +(2\leftrightarrow 3)\nonumber\\
&&+2p_1\inn N\inn p_2 \bigg((\cG(\veps_2^S\inn V\inn\veps_3^S)^{ja_1}+\cG(\veps_2^A\inn V\inn\veps_3^A)^{a_1j})\cI_3+(2\leftrightarrow 3)\bigg)\bigg]\labell{M''2add}
\eeqa
 The above amplitude is invariant under the linear T-duality when the $y$-index is carried by the RR potential, otherwise, it is not invariant. We will consider the T-dual completion of this amplitude in the next section. We have checked that the sum of the amplitudes \reef{M'2}, \reef{M'2add} and \reef{M''2add} does not satisfy the NSNS Ward identity. So the amplitude $A''_2$ is required to make the amplitudes invariant under the NSNS gauge transformations.

Since the amplitude $A''_2$ has the RR factor $(f_1)^{ija_0a_2}p_{2i}p_{3j}$, one has to consider all independent terms with one world volume index $(\cdots)^{a_1}$ which contain one momentum and the two NSNS polarization tensors. In this case there are terms in which the two tensors contract with each other. All possible such terms are
\beqa
\Tr[\veps^S\inn V\inn \veps^S \inn V]\,\,,\,\,\,\Tr[\veps^S\inn N\inn \veps^S \inn N]\,\,,\,\,\,\Tr[\veps^S\inn V\inn \veps^S \inn N]\nonumber\\
\Tr[\veps^A\inn V\inn \veps^A \inn V]\,\,,\,\,\,\Tr[\veps^A\inn N\inn \veps^A \inn N]\,\,,\,\,\,\Tr[\veps^A\inn V\inn \veps^A \inn N]
\eeqa
Since the independent terms must be invariant under the linear T-duality when $a_1\neq y$, we have to consider the combination of the above terms which are invariant under the T-duality. The only possibility is the following combination:
\beqa
\Tr[\veps_2^A\inn V\inn \veps_3^A \inn V]+\Tr[\veps_2^A\inn N\inn \veps_3^A \inn N]-2\Tr[\veps_2^S\inn V\inn \veps_3^S \inn N]
\eeqa
However, the NSNS Ward identity requires other traces as well. The only way that we can make the T-duality invariant combination is to consider dilaton terms as well as the gravitons. Using the T-duality transformation of the dilaton in the string frame, one finds the following combination is invariant under the linear T-duality:
\beqa
\Tr[\veps_2^S\inn V\inn \veps_3^S \inn V]+\Tr[\veps_2^S\inn N\inn \veps_3^S \inn N]-2\Tr[\veps_2^A\inn V\inn \veps_3^A \inn N]+4\Phi _2 \Phi _3\labell{Tr}
\eeqa
where $\Phi$ is the polarization of the dilaton which is one, however, we keep it for clarity. 

Using the above two T-duality invariant combinations, as well as the structures in which the polarization tensors contract with the momentum, one finds the NSNS Ward identity is satisfied provided that the amplitude   $A''_2$ has  the following terms:
\beqa
{ A''}_2&\!\!\!\!\!\!\!\!\!\!\sim\!\!\!\!\!\!\!\!\!\!&\frac{1}{4}(f_1)_{ij}{}^{a_0a_2}p_{2}^ip_{3}^j\bigg[ -\cJ_{15}\cG(p_2\inn N\inn \veps_3^A\inn V\inn \veps_2^A)^{a_1} -\cJ_{15}\cG(p_2\inn V\inn \veps_3^A\inn N\inn \veps_2^A)^{a_1}\nonumber\\
&&+(4\cJ-\cJ_{16}-2\cJ_{5})\cG(p_2\inn V\inn \veps_3^A\inn V\inn \veps_2^A)^{a_1}-(\cJ_{16}-2\cJ_{5})\cG(p_2\inn N\inn \veps_3^A\inn N\inn \veps_2^A)^{a_1}\nonumber\\
&&-2 \bigg(\cJ_4(p_3\inn V\inn \veps_2^S)^{a_1}-\cJ_{12} (p_3\inn N\inn \veps_2^S)^{a_1} +\cJ_3 p_3^{a_1}\Tr[\veps_2^S\inn V]\bigg)\Tr[\veps_3^S\inn V]\nonumber\\
&&+\frac{1}{2}p_3^{a_1}\bigg((2\cJ-\cJ_{16})(\Tr[\veps_2^A\inn N\inn \veps_3^A\inn N]+\Tr[\veps_2^A\inn V\inn \veps_3^A\inn V]-2\Tr[\veps_2^S\inn V\inn \veps_3^S\inn N])\nonumber\\
&&+\cJ_{15}(\Tr[\veps_2^S\inn N\inn \veps_3^S\inn N]+\Tr[\veps_2^S\inn V\inn \veps_3^S\inn V]-2\Tr[\veps_2^A\inn V\inn \veps_3^A\inn N]+4\Phi _2 \Phi _3\bigg) -(2\leftrightarrow 3)\nonumber\\
&&-2\cI_{2}\cG(p_1\inn N\inn \veps_3^A\inn N\inn \veps_2^A)^{a_1}+2\cI_{3}\cG(p_1\inn N\inn \veps_3^A\inn V\inn \veps_2^A)^{a_1}\bigg]\labell{A2addadd}
\eeqa
 This amplitude is also invariant under the linear T-duality when the $y$-index is carried by the RR potential, otherwise, it is not invariant. We will consider the T-dual completion of this amplitude in the next section.  

Note that we have included the dilaton term in above amplitude based on the fact that the amplitude should be consistent with the T-dual Ward identity. As a result the above amplitude is correct in the string frame. However, the direct string theory calculation produces   amplitudes   in the Einstein frame. Therefore, if one is interested in verifying the dilaton amplitude by the direct string theory S-matrix element of one RR and two dilaton vertex operators, one has to transform the S-matrix element to the string frame and then compare with the above result.

The combination of the amplitudes \reef{M'2}, \reef{M'2add}, \reef{M''2add} and \reef{A2addadd}  satisfies the Ward identity corresponding to the NSNS gauge transformations.  However, it does not satisfy the RR Ward identity because the RR factor $(f_1)^{ija_0a_2}$ in them is not the RR field strength. It can easily be extended to the RR invariant amplitudes by extending the RR factor to the RR field strength $(F_1)^{ij a_0a_2}=p_1^{a_0}\veps_1^{ija_2}-p_1^{a_2}\veps_1^{ija_0}+p_1^{i}\veps_1^{a_2ja_0}-p_1^{j}\veps_1^{a_2ia_0}$. The amplitude corresponding to the last two terms does not satisfy the NSNS Ward identity. So one has to add the  amplitude of the  RR $(p+1)$-form with one transverse index. The RR invariance requires the  amplitude to be in terms of RR field strength, \ie $(F_1)^{ia_0a_1a_2}(\cdots)_i$. 
These couplings may   be found by imposing the T-dual Ward identity on the  RR $(p-1)$-form amplitude with structure $(F_1)^{a_0a_1a_3}(\cdots)^{a_2}$ when $a_2=y$. As we have discussed in the previous section, one needs explicit calculation to find the  RR $(p-1)$-form amplitude with structure $(F_1)^{a_0a_1a_3}(\cdots)^{a_2}$.    We leave the details of theses calculations for the future work.

\subsection{RR $(p+3)$-form}

The amplitude for the RR $(p+3)$-form is non-zero when the RR potential has five, four,  and three  transverse indices.  When the RR potential has three transverse indices, the amplitude can be found by applying  the T-dual Ward identity on the amplitude \reef{A2}, \ie
\beqa
{\bf A}_3 &=&{A}_3(f_1)+{\cal A}_3(f_1)\labell{A3}
\eeqa
 where the subscribe 3   refers to the number of the transverse indices of the RR potential. The amplitude  ${\cal A}_3(f_1)$ is the T-dual completion of the amplitude ${\cal A}_2(f_1)$ in \reef{M2}, and ${A}_3(f_1)$ is the T-dual completion of the amplitude $A_2(f_1)$ in \reef{M'2}.  The   amplitude  \reef{M2} does not satisfy the T-dual Ward identity when the $y$-index is carried by the NSNS polarization tensors. The consistency with the T-dual Ward identity requires the following amplitude: 
\beqa
{\cal A}_3&\!\!\!\!\!\sim\!\!\!\!\!& -\frac{1}{4}(f_1)_{ijkl}{}^{a_1}\bigg[(\veps_2^A)^{ij}\bigg(2p_2^kp_2^{a_0}(p_1\inn N\inn\veps_3^S)^{l}\cI_1+(p_3^kp_2^{a_0}\cI_3+p_2^kp_3^{a_0}\cI_2)(p_2\inn N\inn\veps_3^S)^{l}\nonumber\\
&&+(p_2^kp_2^{a_0}\cI_3+p_3^kp_3^{a_0}\cI_2)(p_2\inn V\inn\veps_3^S)^{l}-4p_2^kp_2^{a_0}(p_3\inn V\inn\veps_3^S)^{l}\cI_4\bigg)+(2\leftrightarrow 3)\labell{A''3}\\
&&-\bigg(2p_2^k p_3\inn V\inn p_3 \cI_4+p_3^k(-p_2\inn N\inn p_3\cI_3+p_2\inn V\inn p_3\cI_2)\bigg)\bigg((\veps_2^A)^{ij}(\veps_3^S)^{a_0l}+(2\leftrightarrow 3)\bigg)\bigg]\nonumber
\eeqa
where the RR factor is $(f_1)^{ijkla_1}=p_1^i\veps_1^{jkla_1}-p_1^j\veps_1^{ikla_1}+p_1^k\veps_1^{ijla_1}-p_1^l\veps_1^{ijka_1}$. For simplicity we have considered the amplitude for $p=1$. The above amplitude is invariant under the T-dual Ward identity  when the $y$-index is carried by the RR potential. However, it does not satisfy this identity when the Killing index is carried by the NSNS polarization tensors,  \ie when $a_{0}=y$. We will find the T-dual completion of this amplitude in the next section.

The above amplitude  does not satisfy the Ward identity corresponding to the NSNS  and the RR  gauge transformations. The asymmetry  under the NSNS transformation indicates that the T-dual Ward identity could not capture   terms of the scattering amplitude of the RR $(p+3)$ which have $(f_1 )^{ijkla_1}p_{2i}$ or $(f_1)^{ijkla_1}p_{3i}$. Since all terms in \reef{M2} have either $p_{2i}$ or $p_{3i}$, one finds that the missing terms    should have the factor $(f_1 )^{ijkla_1}p_{2i}p_{3j}$.   Considering all  such terms which have  one   momentum and the two NSNS polarization tensors, with unknown coefficients and imposing the condition that when they combine with the amplitude \reef{A''3} they should satisfy the NSNS Ward identity, one finds the following result:
\beqa
{\cal A'}_3&\!\!\!\!\!\sim\!\!\!\!\!&\frac{1}{4}(f_1 )^{ijkla_1}p_{2i}p_{3j}\bigg[2p_3^{a_0}\cI_4(\veps_2^A)^{kl}\Tr[\veps_3^S\inn V] +\cI_2\bigg(2(p_2\inn N\inn\veps_3^A)^{k}(\veps_2^S)^{a_0l}+ (p_2\inn N\inn\veps_3^S)^{a_0}(\veps_2^A)^{kl}\bigg)\nonumber\\
&&-\cI_3\bigg(2(p_2\inn V\inn\veps_3^A)^{k}(\veps_2^S)^{a_0l}+(p_2\inn V\inn\veps_3^S)^{a_0}(\veps_2^A)^{kl}\bigg)-(2\leftrightarrow 3)+2p_3^{a_0}\cG(\veps_2^S\inn V\inn\veps_3^A)^{kl}\cI_3\nonumber\\
&&+2p_3^{a_0}\cG(\veps_3^S\inn V\inn\veps_2^A)^{kl}\cI_2 - 2\cI_1\bigg(2(p_1\inn N\inn\veps_3^A)^{k}(\veps_2^S)^{a_0l}+(p_1\inn N\inn\veps_3^S)^{a_0}(\veps_2^A)^{kl}\bigg)\bigg]\labell{M3add}
\eeqa
One can verify that the above amplitude is   the T-dual completion of the amplitude \reef{M2add}, as expected. The combination of the amplitudes \reef{A''3} and \reef{M3add} satisfies the NSNS Ward identity, however, they do not satisfy the RR Ward identity. If one includes the amplitude of the RR $(p+3)$-form with four transverse indices which has been found in \cite{Velni:2012sv}, then the RR factor in the above amplitude is extended to the RR field strength $(F_1)^{ijkla_1}=(f_1)^{ijkla_1}+p_1^{a_1}\veps_1^{ijkl}$, \ie the amplitudes ${\cal A}_3(C_{ijk})+{\cal A'}_3(C_{ijk})$ is extended to ${\cal A}_4(F_{ijkl})+{\cal A'}_4(F_{ijkl})$. The amplitude of the RR $(p+3)$-form with four transverse indices has also some terms which become RR gauge invariant after including the amplitude of the RR $(p+3)$-form with five transverse indices \cite{Velni:2012sv}. The RR gauge invariant amplitudes  ${\cal A}_4(F_{ijkl})+{\cal A'}_4(F_{ijkl})$ do not satisfy the NSNS Ward identity which indicates the presence of other amplitude in \reef{A3}.

The amplitude ${A}_3(f_1)$ in \reef{A3} can be found from imposing the invariance of the amplitude \reef{M'2} under the linear T-duality when the Killing index $y$ is carried by the NSNS   polarization tensors in \reef{M'2}. The result is the following:
\beqa
{A}_3&\!\!\!\!\!\sim\!\!\!\!\!&\frac{1}{4}(f_1)_{ijk}{}^{a_0a_1}(\veps_2^A)^{ij}\bigg[2p_2\inn V\inn p_2(p_1\inn N\inn\veps_3^S)^{k}\cI_7-4p_2\inn V\inn p_2(p_3\inn V\inn\veps_3^S)^{k}\cJ_3\nonumber\\
&&+\bigg(p_2\inn N\inn p_3\cJ_{15}+p_2\inn V\inn p_3(\cJ_{16}-2\cJ)\bigg)(p_2\inn N\inn\veps_3^S)^{k}\nonumber\\
&&-\bigg(2p_2\inn N\inn p_3\cJ-p_2\inn V\inn p_2\cJ_1-p_3\inn V\inn p_3\cJ_4\bigg)(p_2\inn V\inn\veps_3^S)^{k}\bigg]+(2\leftrightarrow 3)\labell{A'3}
\eeqa 
where the RR factor is $(f_1)^{ijka_0a_1}=p_1^{a_0}\veps_1^{ijka_1}-p_1^{a_1}\veps_1^{ijka_0}$. Since the NSNS polarization tensors do not carry the world volume index, the above amplitude is invariant under the linear T-duality. However, it does not satisfy the Ward identity corresponding to the NSNS and the RR gauge transformations.

The asymmetry  under the NSNS gauge transformation indicates that the T-dual Ward identity could not capture   terms of the scattering amplitude of the RR $(p+3)$ which have $(f_1 )^{ijka_0a_1}p_{2i}$ or $(f_1)^{ijka_0a_1}p_{3i}$. Since the RR factor is $(f_1)^{ijk}{}^{a_0a_1}$, one finds that there are three types of  missing terms. The first type is the terms which have    $(f_1 )^{ijka_0a_1}p_{2i}$,  the second type is the terms  which  have  $(f_1)^{ijka_0a_1}p_{3i}$, and the third type  is the terms  which  have  $(f_1)^{ijka_0a_1}p_{2i}p_{3j}$. Therefore, the missing terms can be separated as $A'_3=A'_{32}+A'_{33}+A_3''$.  One may consider all  such terms   with unknown coefficients and impose the NSNS Ward identity to find the coefficients.  Alternatively, one may find these amplitudes   by imposing the T-dual Ward identity on the amplitudes \reef{M'2add},  \reef{M''2add} and \reef{A2addadd}. We are going to preform  the latter calculations.  The T-dual completion of the amplitude \reef{M'2add} which has  the terms of the first type, is the following:
\beqa
{A'}_{32} &\!\!\!\!\!\sim\!\!\!\!\!&\frac{1}{4}(f_1 )_{ijk}{}^{a_0a_1}p_{2}^i\bigg[(p_3\inn V\inn\veps_2^S)^{k}\bigg((p_2\inn N\inn\veps_3^A)^{j}\cJ_{15}+(p_2\inn V\inn\veps_3^A)^{j}\,(\cJ_{16}-4\cJ+2\cJ_5)\bigg)\nonumber\\
 &&+(p_3\inn N\inn\veps_2^S)^{k}\bigg(2 \cI_{2}(p_1\inn N\inn\veps_3^A)^{j}+(p_2\inn N\inn\veps_3^A)^{j}\,(\cJ_{16}-2\cJ_5)+(p_2\inn V\inn\veps_3^A)^{j}\cJ_{15}\bigg)\nonumber\\
 &&+\frac{1}{2}(\veps_2^A)^{jk}\bigg(p_2\inn N\inn\veps_3^S\inn N\inn p_2(\cJ_{16}-2\cJ_5)-4\cJ_{4} p_3\inn V\inn\veps_3^S\inn N\inn p_2+4\cI_2p_2\inn N\inn\veps_3^S\inn N\inn p_1\nonumber\\
&&+p_2\inn V\inn\veps_3^S\inn V\inn p_2(\cJ_{16}-4\cJ+2\cJ_5)+4p_3\inn V\inn\veps_3^S\inn V\inn p_2\cJ_{12}-2 p_2\inn V\inn\veps_3^S\inn N\inn p_1\cI_3\nonumber\\
&&+2p_2\inn V\inn\veps_3^S\inn N\inn p_2\cJ_{15}+(-2\cI_{4}p_1\inn N\inn p_3+\cJ_4p_2\inn N\inn p_3-\cJ_{12}p_2\inn V\inn p_3)\Tr[\veps_3^S\inn V]\bigg)\nonumber\\
&&-2p_3\inn V\inn p_3\cG(\veps_2^S\inn V\inn\veps_3^A)^{jk}\cJ_{12}\bigg]+(2\leftrightarrow 3)\labell{A32}
\eeqa
Since the NSNS polarization tensors do not carry the world volume index, the above amplitude is invariant under the linear T-duality.

The T-dual completion of the amplitude \reef{M''2add} which has  the terms of the second type, is the following:
\beqa
{A'}_{33}&\!\!\!\!\!\sim\!\!\!\!\!& \frac{1}{4}(f_1 )_{ijk}{}^{a_0a_1} p_{3}^i\bigg[-2p_2\inn V\inn p_2  \cG(\veps_2^S\inn V\inn\veps_3^A)^{j k}\cJ_1+(p_3\inn V\inn\veps_2^S)^{k}\bigg(2(p_1\inn N\inn\veps_3^A)^{j}\cI_2\nonumber\\
&&+(p_2\inn N\inn\veps_3^A)^{j}(\cJ_{16}-4\cJ-2\cJ_5)+(p_2\inn V\inn \veps_3^A)^{j}\cJ_{15}\bigg)\nonumber\\
&&+2(p_1\inn N\inn\veps_2^S)^{k}\bigg(2 (p_1\inn N\inn\veps_3^A)^{j}\cI_{1}-(p_2\inn N\inn\veps_3^A)^{j}\cI_{2}+(p_2\inn V\inn\veps_3^A)^{j}\cI_{3}\bigg)\nonumber\\
&&-4(p_2\inn V\inn\veps_2^S)^{k}\bigg(2 (p_1\inn N\inn\veps_3^A)^{j}\cI_{7}-(p_2\inn N\inn\veps_3^A)^{j}\cJ_{2}+(p_2\inn V\inn\veps_3^A)^{j}\cJ_{1}\bigg)\nonumber\\
&&+(p_3\inn N\inn\veps_2^S)^{k}\bigg((p_2\inn N\inn\veps_3^A)^{j}\cJ_{15}+(p_2\inn V\inn\veps_3^A)^{j}(\cJ_{16}+2\cJ_5)\bigg)\nonumber\\
&&+\frac{1}{2}(\veps_2^A)^{jk}\bigg(4(p_1\inn N\inn p_2\cI_{4}-p_2\inn V\inn p_2\cJ_{3})\Tr[\veps_3^S\inn V]+2p_2\inn V\inn\veps_3^S\inn N\inn p_2\,(\cJ_{16}-2\cJ)\nonumber\\
&&+(p_2\inn N\inn\veps_3^S\inn N\inn p_2\,+p_2\inn V\inn\veps_3^S\inn V\inn p_2)\cJ_{15}+2p_2\inn V\inn\veps_3^S\inn N\inn p_1\, \cI_2\bigg) +(2\leftrightarrow 3)\nonumber\\
&&+2p_1\inn N\inn p_2p_{3}^i\bigg(\cG(\veps_2^S\inn V\inn\veps_3^A)^{jk}\cI_3+(2\leftrightarrow 3)\bigg)\bigg]\labell{A33}
\eeqa
The NSNS polarization tensors do not carry the world volume index, so the above amplitude is also invariant under the linear T-duality. Wile the first $(2\leftrightarrow 3)$ means the interchange of the labels $2, 3$ for all expressions from the beginning up to that point, including the overall factor, the second $(2\leftrightarrow 3)$ means the interchange of the labels $2, 3$ only for the term in the parenthesis in the last line.

The T-dual completion of the amplitude \reef{A2addadd} which has  the terms of the third type, is the following:
\beqa
{ A''}_3&\!\!\!\!\!\sim\!\!\!\!\!&\frac{1}{4} (f_1)^{ijka_0a_1}p_{2i}p_{3j}\bigg[-2\bigg(\cJ_4 (p_3\inn V\inn \veps_2^A)_k -\cJ_{12} (p_3\inn N\inn \veps_2^A)_k\bigg)\Tr[\veps_3^S\inn V] \nonumber\\
&&-\cJ_{15}\cG(p_2\inn V\inn \veps_3^A\inn N\inn \veps_2^S)_k+(4\cJ-\cJ_{16}-2\cJ_{5})\cG(p_2\inn V\inn \veps_3^A\inn V\inn \veps_2^S)_k\nonumber\\
&&-(\cJ_{16}-2\cJ_{5})\cG(p_2\inn N\inn \veps_3^A\inn N\inn \veps_2^S)_k -\cJ_{15}\cG(p_2\inn N\inn \veps_3^A\inn V\inn \veps_2^S)_k-(2\leftrightarrow 3)\nonumber\\
&& -2\cI_{2}\cG(p_1\inn N\inn \veps_3^A\inn N\inn \veps_2^S)_k+2\cI_{3}\cG(p_1\inn N\inn \veps_3^A\inn V\inn \veps_2^S)_k\bigg]\labell{A3addadd}
\eeqa
This amplitude is also invariant under the linear T-duality. The sum of the amplitudes \reef{A'3}, \reef{A32}, \reef{A33} and \reef{A3addadd} satisfies the Ward identity corresponding to the NSNS gauge transformations. However, it does not satisfy the RR Ward identity because the RR factor $(f_1)^{ijka_0a_1}$ is not the RR field strength. It can easily be extended to the RR invariant amplitude by extending the RR factor to the RR field strength $(F_1)^{ijk a_0a_1}=(f_1)^{ijka_0a_1}+p_1^{i}\veps_1^{a_1jka_0}-p_1^{j}\veps_1^{a_1ika_0}+p_1^{k}\veps_1^{a_1ija_0}$. In this case, the amplitudes corresponding to the last three terms   satisfy the NSNS Ward identity.

\subsection{RR $(p+5)$-form}

The amplitude for the RR $(p+5)$-form is non-zero when the RR potential has six, five, and four   transverse indices.  When the RR potential has four transverse indices, the amplitude can be found by applying  the T-dual Ward identity on the amplitude \reef{A3}, \ie
\beqa
{\bf A}_4 &=&{\cal A}_4(f_1)\labell{A41}
\eeqa
 where the subscribe 4   refers to the number of the transverse indices of the RR potential. The amplitude  ${\cal A}_4(f_1)$ which is the T-dual completion of the amplitude ${\cal A}_3(f_1)$ in \reef{A''3} is 
 \beqa
{\cal A}_4 & \!\!\!\sim\!\!\! & \frac{1}{8}(f_1)_{ijklm}{}^{a_0}\bigg(2p_2^kp_3\inn V\inn p_3\cI_4+p_3^k(\cI_2p_2\inn V\inn p_3-\cI_3p_2\inn N\inn p_3)\bigg)(\veps_2^A)^{ij}(\veps_3^A)^{lm}\labell{A4}
\eeqa 
where the RR factor is $(f_1)^{ijklma_0}=p_1^{[i}\veps_1^{jklm]a_0}$. Since the NSNS polarization tensors do not carry the world volume index, the above amplitude is invariant under the linear T-duality. However, it does not satisfy the  Ward identity corresponding to the NSNS and RR gauge transformations.

The asymmetry  under the NSNS transformation indicates that the T-dual Ward identity could not capture   terms of the scattering amplitude of the RR $(p+5)$ which have $(f_1 )^{ijklma_0}p_{2i}$ or $(f_1)^{ijklma_0}p_{3i}$. Since all terms in \reef{A''3} have either $p_{2i}$ or $p_{3i}$, one finds that the missing terms  corresponding to the above amplitude  should have the factor $(f_1 )^{ijklma_0}p_{2i}p_{3j}$.   Considering all  such terms which have  one   momentum and the two NSNS polarization tensors, with unknown coefficients and imposing the condition that when they combine with the amplitude \reef{A4} they should satisfy the NSNS Ward identity, one finds the following result:
\beqa
{\cal A'}_4&\!\!\!\!\!\sim\!\!\!\!\!&  \frac{1}{4}(f_1)_{ijklm}{}^{a_0} p_2^ip_3^j(\veps_2^A)^{lm}\bigg[(p_2\inn N \inn \veps_3^A)^k\cI_2-(p_2\inn V \inn \veps_3^A)^k \cI_3\bigg]+(2\leftrightarrow 3)\nonumber\\
&&-\frac{1}{2}(f_1)_{ijklm}{}^{a_0}p_2^ip_3^j(\veps_2^A)^{lm}(p_1\inn N \inn \veps_3^A)^k\cI_1 \labell{A4add}
\eeqa
The above amplitude is also the T-dual completion of the amplitude \reef{M3add}.   There is no contraction between the NSNS polarization tensors and the world volume form, so this amplitude, like \reef{A4}, is   invariant under the linear T-duality. 

The combination of the above two amplitudes satisfy the NSNS Ward identity, however, they do not satisfy the RR Ward identity. To extend the amplitude ${\cal A}_4+{\cal A'}_4$ to satisfy the RR Ward identity, one has to extend the RR factor to the RR field strength $(F_1)^{ijklma_0}=(f_1)^{ijklma_0}-p_1^{a_0}\veps_1^{ijklm}$, \ie the amplitudes ${\cal A}_4(C_{ijkl})+{\cal A'}_4(C_{ijkl})$ is extended to ${\cal A}_5(F_{ijklm})+{\cal A'}_5(F_{ijklm})$. This can be done by including  the amplitude of the RR $(p+5)$-form with five transverse indices which has been found in \cite{Velni:2012sv}. The amplitude of the RR $(p+5)$-form with five transverse indices has also some terms which become RR gauge invariant after including the amplitude of the RR $(p+5)$-form with six transverse indices \cite{Velni:2012sv}. These amplitudes and the amplitude ${\cal A}_5(F_{ijklm})+{\cal A'}_5(F_{ijklm})$ are exactly equal to the amplitudes that has been calculated explicitly in string theory for the case that the RR potential is $(p+5)$-form \cite{Garousi:2011ut}. So they satisfies the NSNS Ward identity as well as the RR Ward identity.

Therefore, the S-matrix elements of one RR and two NSNS can be classified into the following multiplets in terms of the RR field strength: One  T-dual multiplet  which has been found in \cite{Velni:2012sv} (see eq.(15) in \cite{Velni:2012sv}) has the following structure:
\beqa
 {A}_2(F_{ij}^{(p-2)})\rightarrow {A}_3(F_{ijk}^{(p)})\rightarrow {A}_4(F_{ijkl}^{(p+2)})\rightarrow {A}_5(F_{ijklm}^{(p+4)})\rightarrow {A}_6(F_{ijklmn}^{(p+6)})\labell{Ai1}
 \eeqa
The T-dual multiplet satisfies the RR Ward identity, however, it does not satisfy the NSNS Ward identity. Another multiplet   has the following structure:
\beqa
\matrix{\cA_1(F_i ^{(p-2)})& \!\!\!\!\!\rightarrow \!\!\!\!\!& \cA_2(F_{ij} ^{(p)})& \!\!\!\!\!\rightarrow \!\!\!\!\!&\cA_3(F_{ijk} ^{(p+2)})&  \!\!\!\!\!\rightarrow \!\!\!\!\! & \cA_4(F_{ijkl} ^{(p+4)})& \!\!\!\!\!\rightarrow \!\!\!\!\!&\cA_5(F_{ijklm} ^{(p+6)})\cr 
&&\downarrow & &\downarrow&& \downarrow&& \downarrow\cr
&&  \cA'_2(F_{ij} ^{(p)})& \!\!\!\!\!\rightarrow \!\!\!\!\!&\cA'_3(F_{ijk} ^{(p+2)})& \!\!\!\!\!\rightarrow \!\!\!\!\!&\cA'_4(F_{ijkl} ^{(p+4)})& \!\!\!\!\!\rightarrow \!\!\!\!\!&\cA'_5 (F_{ijklm} ^{(p+6)})}
 \labell{close31} 
 \eeqa 
where $\cA_1$ is the amplitude \reef{A''0A}, the amplitudes $\cA_2+\cA'_2$ are the amplitudes \reef{M1} and \reef{M1add}, the amplitudes $\cA_3+\cA'_3$ are the amplitudes \reef{M2} and \reef{M2add}, the amplitudes $\cA_4+\cA'_4$ are the amplitudes \reef{A''3} and \reef{M3add}, and the amplitudes $\cA_5+\cA'_5$ are the amplitudes \reef{A4} and \reef{A4add} in which the RR factor $f_1$ is replaced by the RR field strength $F_1$. The third multiplet  has the following structure:
\beqa
\matrix{A_0(F ^{(p-2)})& \!\!\!\!\!\rightarrow \!\!\!\!\!& A_1(F_i ^{(p)})& \!\!\!\!\!\rightarrow \!\!\!\!\!&A_2(F_{ij} ^{(p+2)})&  \!\!\!\!\!\rightarrow \!\!\!\!\! & A_3(F_{ijk} ^{(p+4)})\cr 
&&\downarrow & &\downarrow&& \downarrow\cr
&&  A'_1(F_i ^{(p)})& \!\!\!\!\!\rightarrow \!\!\!\!\!&A'_2(F_{ij} ^{(p+2)})& \!\!\!\!\!\rightarrow \!\!\!\!\!&A'_3(F_{ijk} ^{(p+4)}) }\labell{3close}
   \eeqa  
where  $A_0$ is the amplitude \reef{A'0A}. The amplitudes $A_1+A'_1$ are the amplitudes \reef{A'1}, \reef{M'1add} and \reef{M''1add}, the  amplitudes $A_2+A'_2$ are the amplitudes \reef{M'2}, \reef{M'2add}, \reef{M''2add} and \reef{A2addadd},  and the  amplitudes $A_3+A'_3$ are the amplitudes \reef{A'3}, \reef{A32}, \reef{A33} and \reef{A3addadd} in which the RR factor $f_1$ is replaced by the RR field strength $F_1$. The last multiplet would have the following structure: 
\beqa
\matrix{A_0(F ^{(p)})& \!\!\!\!\!\rightarrow \!\!\!\!\!& A_1(F_{ i}^{(p+2)})  \cr
&& \downarrow\cr
&&  {A_1'} (F_{ i}^{(p+2)})\cr
} \labell{Fiii1} 
\eeqa 
 The first component of the above multiplet may be found from the explicit string theory calculation in which we are not interested in this paper. Using the T-dual Ward identity on the first component,  the second component then would be easily found, as we have done for many other cases in this paper.
	
	\section{Discussion}
	
	In this paper we have used the constraints that the S-matrix elements should satisfy the Ward identity corresponding to the gauge symmetries and the T-duality, to find the D$_p$-brane world volume amplitude of various RR $n$-forms from the known amplitudes of the RR $(p-3)$-form.  Using this constraint, we have found various S-matrix elements  of one RR, one NSNS and one NS states,  and the S-matrix elements of one RR and two NSNS states. 
	
	We have found that the  Ward identities corresponding to the  combination of the T-duality and the gauge transformations,  are powerful enough to find all the S-matrix multiplets which are connected by these Ward identities. However, the Ward identities corresponding to  the  gauge transformations along, are not powerful enough to find all the amplitudes which are connected   by these Ward identities.  For the case of two closed and one open strings, the T-dual multiplets are \reef{Fi}, \reef{Fii}, \reef{Fiii} , and for the case of three closed strings ,  the T-dual multiplets are \reef{Ai1}, \reef{close31}, \reef{3close}, \reef{Fiii1}. 
	
	In each multiplet,  the different components are connected by the T-dual and the NSNS Ward identities.   On the other hand, the  components of all the T-dual multiplets which have a specific RR field strength, are connected by the NSNS/NS Ward identity, \eg the $F^{(p)}$-component in the multiplets \reef{Fi}, \reef{Fii}, \reef{Fiii}, and the $F^{(p)}$-component in the multiplets \reef{Ai1}, \reef{close31}, \reef{3close}, \reef{Fiii1}  should satisfy the Ward identities corresponding to the gauge transformations. In the former case, the amplitude $A_0(F^{(p)})$ has been found   by these Ward identities, \ie \reef{Fp}, however, there are two integrals and one constraint. The explicit form of the integrals can be found only by direct calculation of the corresponding S-matrix element in the string theory. In the latter case,   the Ward identities produce   many new integrals and constraint equations. It would be interesting to find this amplitude by the explicit string theory calculations, and then find its corresponding multiplet \reef{Fiii1}. It would be also interesting to confirm  the amplitudes that we have found in this paper by explicit string theory calculations.
	
	The S-matrix elements of three closed strings that   have been found  in this paper can be analyzed  at low energy to extract the appropriate couplings of one RR and two NSNS states in the field theory at order $\alpha'^2$. In performing this calculation, one needs the $\alpha'$-expansion of  the integrals that appear in the amplitudes. The  $\alpha'$-expansion of the integrals $\cI_1,\, \cI_2,\, \cI_7$ and $\cJ,\, \cJ_1,\, \cJ_2,\, \cJ_3,\,\cJ_5,\, \cJ_{13},\cJ_{14}$  have been found in  \cite{Garousi:2010bm,Garousi:2011ut} for the special kinematic setup where $p_2 \inn D\inn p_3=0$ and $p_2 \inn p_3=0$. The above integrals are similar to the integrals $I_0,\cdots, I_{10}$ that have been found in \cite{Becker:2011ar}.  The relation between the two set of integrals is  
\beqa
&&\cI_1=I_{10},\,\,\,\,\cI_2=I_5-I_9,\,\,\,\ \cI_7=-\frac{1}{2} I_4,\,\,\,\ \cJ=2 I_0,\,\ \cJ_1=-(I_6+I_7) ,\nonumber\\
&& \cJ_2=-2I_0 -I_8+I_{10},\,\,\,\cJ_3=-I_3,\,\,\,\  \cJ_5=I_8,\,\,\,\  \cJ_{13}=2I_0+I_2,\,\,\,\  \cJ_{14}=2I_0+I_1\nonumber
\eeqa 
The low energy expansion of the integrals $I_0,\cdots, I_{10}$, for the general setup, have been found in  \cite{Becker:2011ar}.  Using them, one can find the $\alpha'$-expansion of the amplitudes which contain various massless poles as well as contact terms. To find the couplings of one RR and two NSNS states at order $\alpha'^2$,  one has to first calculate the massless   poles in field theory   and  then subtract them from the massless poles of the string theory amplitude.   The massless poles at order $\alpha'^2$ are simple closed string poles, simple open string poles and double open string poles. The closed string poles should be  reproduced by the supergravity and the brane couplings of two closed strings at order $\alpha'^2$ \cite{Bachas:1999um,Garousi:2010ki,Garousi:2011fc}. The simple open string poles should be  reproduced by the DBI or CS action and the brane couplings of two closed and one open strings at order $\alpha'^2$ which can be found from the amplitudes in section 3.  The double open string poles should be reproduced by the DBI or CS action and the brane couplings of one closed and two open string at order $\alpha'^2$ \cite{Hashimoto:1996kf,Garousi:1998fg}. 

The subtraction of field theory massless poles from the string theory amplitude may   add some   extra contact terms to the contact terms of the string theory amplitude. 
For the amplitudes which involve only the antisymmetric NSNS states, one may expect  the extra contact terms to be avoided by  writing both the string theory amplitude and the  field theory massless poles,  in terms of B-field strength $H$.    This can be done based on the fact that the S-matrix elements must satisfy the Ward identity corresponding to the B-field gauge transformation. In the field theory side,  the bulk couplings are in terms of   $H$, and the brane couplings are either in terms of $H$ or in terms of $\tilde{B}=B+2\pi\alpha'\cF$. As a result,  one can  calculate the massless   poles in the field theory   in terms of $H$. In fact, the open string poles of the scattering amplitude in which the gauge boson part of $\tilde{B}$ propagates, can be combined with the contact terms resulting from the $B$-field part of $\tilde{B}$ to  write the amplitude in terms of $H$ \cite{Garousi:2010bm,Garousi:2011ut}. While the field theory massless poles can be calculated uniquely in terms of $H$,  there is no unique way, in general,  to write the string theory amplitude in terms of $H$. 

For the case of the RR $(p-3)$-form that has been studied in \cite{Garousi:2011ut}, there is a unique way to write the string theory amplitude in terms of $H$.  Hence, in that case, one does not need to calculate the field theory massless poles. The contact terms of the string theory in terms of $H$ gives the correct couplings in field theory. For the case of the RR $(p+1)$-form, we have checked that there is no unique way to write the amplitude in terms of $H$. Therefore, even in this, one has to calculate the massless poles and subtract them from the string theory amplitude to find the contact terms. After finding all contact terms, one should be able to write them in terms of the field strengths of the external states. We leave the details of this calculation for the future works.

	{\bf Acknowledgment}: K.B.V would like to thank A. Jalali  for checking some of the amplitudes and D. Mahdavian Yekta for useful discussions. This work is supported by Ferdowsi University of Mashhad under  grant 3/19896-1390/09/27.

\end{document}